\documentclass[oneside,11pt]{article}

\usepackage[utf8]{inputenc}
\usepackage[T1]{fontenc}

\usepackage{amsthm}
\usepackage{amsmath}
\usepackage{amsfonts}
\usepackage{amssymb}

\usepackage{graphicx} 
\usepackage{float}
\usepackage{booktabs}
\usepackage{multirow}
\usepackage{rotating}
\usepackage{array}
\usepackage{xcolor}
\usepackage{subfig}
\usepackage[ruled]{algorithm2e}

\newcolumntype{R}[1]{>{\raggedleft\let\newline\\\arraybackslash\hspace{0pt}}m{#1}}
\newcolumntype{L}[1]{>{\raggedright\let\newline\\\arraybackslash\hspace{0pt}}m{#1}}

\usepackage{breakcites}

\usepackage{nowidow}

\usepackage{enumitem}

\usepackage[top=1.5cm,bottom=2cm,right=2.5cm,left=2.5cm]{geometry}
\usepackage{titling}

\usepackage{natbib}

\usepackage{ifplatform}

\usepackage{textcomp}

\ifwindows
\fi

\allowdisplaybreaks

\newcommand{\R}{\mathbb{R}}

\newcommand{\Z}{\mathbb{Z}}

\DeclareFontFamily{OT1}{pzc}{}
\DeclareFontShape{OT1}{pzc}{m}{it}{<-> s * [1.10] pzcmi7t}{}
\DeclareMathAlphabet{\mathpzc}{OT1}{pzc}{m}{it}

\newtheorem{proposition}{Proposition}
\newtheorem{lemma}{Lemma}
\newtheorem{algo}{Algorithm}

\renewenvironment{proof}[1]{\textit{Proof#1.}}{\qed\\} 

\newcommand{\defin}{:=}
\newcommand{\E}{\mathbb{E}}
\newcommand{\n}{^{(n)}}
\newcommand{\ny}{n\rightarrow\infty}
\newcommand{\Sigb}{{\boldsymbol \Sigma}}
\newcommand{\Gamb}{{\boldsymbol \Gamma}}

\usepackage[pdftex,final,bookmarksnumbered,bookmarksopen=false,breaklinks,colorlinks]{hyperref}
\hypersetup{
	final,
	bookmarksnumbered=true,
	bookmarksopen=true,
	bookmarksopenlevel=0,
	unicode=false,          
	pdftoolbar=true,        
	pdfmenubar=true,        
	pdffitwindow=false,     
	pdftitle={Nonparametric tests of independence for circular data based on trigonometric moments},    
	pdfdisplaydoctitle=true, 
	pdftoolbar=true,
	pdfmenubar=true,
	pdflang={English},
	pdfauthor={Eduardo Garcia-Portugues, Pierre Lafaye de Micheaux, Simos G. Meintanis, Thomas Verdebout},     
	pdfsubject={arXiv paper},   
	pdfcreator={Eduardo Garcia-Portugues},   
	pdfproducer={Eduardo Garcia-Portugues}, 
	pdfkeywords={Characteristic function}{Circular data}{Directional data}{Independence}{Trigonometric moments}, 
	pdfnewwindow=true,      
	breaklinks=true,
	hidelinks,       
	linkcolor=black,          
	citecolor=black,        
	filecolor=magenta,      
	urlcolor=cyan           
}

\newif\ifmain
\maintrue
\newif\ifsupplement
\supplementtrue

\newif\iffigstabs
\figstabstrue

\begin{document}

\ifmain

\title{Nonparametric tests of independence for circular data based on trigonometric moments}
\setlength{\droptitle}{-1cm}
\predate{}%
\postdate{}%
\date{}

\author{Eduardo Garc\'ia-Portugu\'es$^{1,8}$, Pierre Lafaye de Micheaux$^{2,3}$,\\
	 Simos G. Meintanis$^{4,5}$, and Thomas Verdebout$^{6,7}$}
\footnotetext[1]{Department of Statistics, Carlos III University of Madrid (Spain).}
\footnotetext[2]{School of Mathematics and Statistics, University of New South Wales (Astralia).}
\footnotetext[3]{Institut Desbrest d'Epid\'{e}miologie et de Sant\'{e} Publique, Universit\'e de Montpellier (France).}
\footnotetext[4]{Department of Economics, National and Kapodistrian University of Athens (Greece).}
\footnotetext[5]{Unit for Pure and Applied Analytics, North-West University (South Africa).}
\footnotetext[6]{D\'{e}partement de Math\'{e}matique, Universit\'{e} libre de Bruxelles (Belgium).}
\footnotetext[7]{ECARES, Universit\'{e} libre de Bruxelles (Belgium).}
\footnotetext[8]{Corresponding author. e-mail: \href{mailto:edgarcia@est-econ.uc3m.es}{edgarcia@est-econ.uc3m.es}.}
\maketitle

\begin{abstract}
	We introduce nonparametric tests of independence for bivariate circular data based on trigonometric moments. Our contributions lie in (\textit{i}) proposing nonparametric tests that are locally and asymptotically optimal against bivariate cosine von Mises alternatives and (\textit{ii}) extending these tests, via the empirical characteristic function, to obtain consistent tests against broader sets of alternatives, eventually being omnibus. We thus provide a collection of trigonometric-based tests of varying generality and known optimalities. The large-sample behaviours of the tests under the null and alternative hypotheses are obtained, while simulations show that the new tests are competitive against previous proposals. Two data applications in astronomy and forest science illustrate the usage of the tests.
\end{abstract}
\begin{flushleft}
\end{flushleft}

\section{Introduction}
\label{sec:intro}

The goal of this paper is to introduce new tests of independence between two circular random variables $\vartheta^{(1)}$ and $\vartheta^{(2)}$ that are supported on $\mathbb{T}\defin[-\pi,\pi)$. Given an independent and identically distributed sample $\big(\vartheta^{(1)}_1,\vartheta^{(2)}_1\big),\ldots,\big(\vartheta^{(1)}_n,\vartheta^{(2)}_n\big)$, we wish to test the null hypothesis ${\cal{H}}_0$ of independence between $\vartheta^{(1)}$ and $\vartheta^{(2)}$, against the general alternative ${\cal H}_1$ consisting on the negation of ${\cal H}_0$. This fundamental testing problem has relevant applications in fields where circular data are common, such as in astronomy, biology, geology, and forest science, to name just a few.\\

A variety of tailored statistical methods for the analysis of data comprised by directions, such as circular data, have been developed in the last decades; see the general treatments of \cite{Mardia1999a}, \cite{Jammalamadaka2001}, and \cite{Ley2017a}, as well as the recent review of \cite{Pewsey2021}. In particular, the analysis of data on $\mathbb{T}^2$ that is generated by a pair of angular variables, referred to as `circular-circular' or `toroidal' data, has attracted a sizeable number of modelling proposals in the recent years \cite[Section 3.2]{Pewsey2021}. This interest has been boosted by applications in bioinformatics, where a sequence of dihedral angles characterizes a protein's three-dimensional backbone \citep[e.g.,][]{Boomsma2008}. In addition, the development of toroidal distributions is intimately related with the design of models for circular time series \citep{Wehrly1980} that naturally appear in a variety of other fields such as astronomy and forest science; see Section \ref{sec:data}.\\

Much of the modelling effort for toroidal data has been dominated by the search for bivariate extensions of the von Mises distribution, often regarded as the `circular Gaussian' distribution. The first of such proposals was the bivariate von Mises density of \cite{Mardia1975b}, considered as an overparametrized model due to its eight parameters. This motivated the six-parameter submodel of \cite{Rivest1988} and the five-parameter `sine' \citep{Singh2002}, `cosine' \citep{Mardia2007}, and `hybrid' \citep{Kent2008} submodels. The properties of the last three were compared in \cite{Kent2008} and \cite{Mardia2012}. A different modelling pathway was initiated with the family of copula-structured toroidal densities by \cite{Wehrly1980}, whose most successful representative is the bivariate wrapped Cauchy distribution \citep{Kato2015a}. \\

Investigating relationships between variables is central to many scientific studies, and tests of independence typically precede any attempt at modelling association. Consequently, many contributions in directional statistics have been dealing with correlation, dependence, and tests for independence. Measures of circular correlation have been put forward by \cite{Watson1967a}, \cite{Jupp1980}, \cite{Shieh1994}, and more recently by \cite{Zhan2019}. In a different direction, \cite{Rothman1971} introduced a version of the Cram\'er--von Mises test of independence. In parametric contexts related with the models of the previous paragraph, one may resort to the likelihood-based tests suggested by \cite{Mardia1978}, \cite{Puri1977}, and \cite{Shieh2005}. Finally, for testing independence in data with mixed directional/linear components, smoothing-based tests have been proposed by \cite{Garcia-Portugues2015}. \\

When testing independence, nonparametric methods based on the characteristic function have also been employed as alternatives to non-omnibus tests based on association coefficients and to smoothing-based tests that exhibit the familiar drawbacks of bandwidth selection and slow convergence. These tests exploit the factorization characterization of the joint characteristic function of independent random variables. This property propagated `Fourier'-type tests in the past, going back as far as \cite{CH82} and \cite{Csorgo1985}. Since then, Fourier methods have enjoyed increasing popularity, finally reaching some sort of climax with the introduction of the novel notion of `distance correlation' \citep{Szekely2007}, and beyond. Indicatively, we refer to the contributions by \cite{Gretton2005}, \cite{Szekely2007}, \cite{Meintanis2008}, \cite{Hlavka2011}, \cite{Fan2017}, \cite{CMZ}, and \cite{CZ19}, all of which propose tests of independence in varying settings and different levels of generality, but always with the characteristic function being the underlying notion. This popularity notwithstanding, and despite the fact that testing based on characteristic functions is not unfamiliar to circular data \citep{Meintanis2019}, the use of characteristic functions for testing independence of non-linear data remains substantially unexplored.\\

We introduce in this paper nonparametric tests of independence for toroidal data based on trigonometric moments. We first propose nonparametric tests using joint cosine moments that are locally and asymptotically optimal against sequences of bivariate cosine von Mises alternatives, and for which the powers of the tests are explicitly obtained. We then extend these tests, via the empirical characteristic function, to more general multiple-orders tests that merge cosine and sine moments, and that are consistent against broader sets of alternatives. We obtain usable asymptotic null distributions for all the test statistics, thus avoiding their calibration by resampling methods. We then propose a characteristic function-based omnibus test with a tractable computational form that can be efficiently calibrated using permutations. Simulations corroborate the adequate finite-sample null and non-null behaviour of the tests, as well as their competitiveness against other testing approaches based on association coefficients. Two data applications are provided, one on the study on the temporal dependence of long-period comet records and another in the evaluation of the dependence between the orientations of Portuguese wildfires.

\section{A cosine test of independence}
\label{sec:nonparamtests}

\subsection{Genesis and null asymptotic distribution}

Our objective is to test the null hypothesis ${\cal H}_0$ of independence between $\vartheta^{(1)}$ and $\vartheta^{(2)}$. Without loss of generality (see Proposition \ref{invariance} below), we assume that $\vartheta^{(j)}$ is circularly centred, i.e., such that its circular mean $\mu^{(j)}\defin\mathrm{atan2}\big(\E\big[\sin\big(\vartheta^{(j)}\big)\big],\E\big[\cos\big(\vartheta^{(j)}\big)\big]\big)$ is zero, $j=1,2$, where $\mathrm{atan2}(y,x)\in\mathbb{T}$ is the argument of the complex number $x+\mathrm{i}y$. Given an independent and identically distributed sample  $\big(\vartheta_1^{(1)}, \vartheta_1^{(2)}\big), \ldots, \big(\vartheta_n^{(1)}, \vartheta_n^{(2)}\big)$ from $(\vartheta^{(1)},\vartheta^{(2)})$, we consider the empirical versions 
\begin{align*}
	\hat{\cal J}_{jc}(r)&\defin n^{-1} \sum_{i=1}^n \cos\big(r \vartheta_i^{(j)}\big),\quad \hat{\cal J}_{js}(r)\defin n^{-1} \sum_{i=1}^n \sin \big(r \vartheta_i^{(j)}\big),\quad j=1,2,\\
	\hat{\cal J}_{c}(r_1,r_2)&\defin n^{-1} \sum_{i=1}^n \cos \big(r_1\vartheta_i^{(1)}+r_2\vartheta_i^{(2)}\big),\quad
	\hat{\cal J}_{s}(r_1,r_2)\defin n^{-1} \sum_{i=1}^n \sin \big(r_1\vartheta_i^{(1)}+r_2\vartheta_i^{(2)}\big),
\end{align*}
of the respective marginal `cosine' and `sine' population moments (as well as their `addition' forms) given by
\begin{align*}
	{\cal J}_{jc}(r) &\defin \E\big[\cos\big(r\vartheta^{(j)}\big)\big],\quad
	{\cal J}_{js}(r)\defin\E\big[\sin\big(r\vartheta^{(j)}\big)\big],\quad j=1,2,\\
	{\cal J}_{c}(r_1,r_2) &\defin \E\big[\cos \big(r_1\vartheta^{(1)}+r_2\vartheta^{(2)}\big)\big],\quad {\cal J}_{s}(r_1,r_2) \defin \E\big[\sin \big(r_1\vartheta^{(1)}+r_2\vartheta^{(2)}\big)\big].
\end{align*}
Here $r$, $r_1$, and $r_2$ are reals, although we will soon restrict to integer numbers; see below \eqref{null1}.\\

Based on the form of the `cosine addition moment', we have that, under the null the hypothesis of independence,
\begin{align}\label{JJ}
	{\cal J}_{c}(r_1,r_2) = {\cal J}_{1c}(r_1) {\cal J}_{2c} (r_2)-{\cal J}_{1s}(r_1){\cal J}_{2s}(r_2).
\end{align}
Based on \eqref{JJ}, it is very natural to consider tests that reject ${\cal H}_0$ for large absolute values of the statistic
\begin{align} \label{realpart}
	D_{c}\n(r_1, r_2)\defin\hat{\cal J}_{c}(r_1,r_2)- \hat{\cal J}_{1c}(r_1) \hat{\cal J}_{2c}(r_2) + \hat{\cal J}_{1s}(r_1)\hat{\cal J}_{2s}(r_2),
\end{align}
since, for any $(r_1,r_2) \in \R^2$, $D_{c}\n(r_1, r_2)$ will be close to zero under ${\cal H}_0$. The following proposition provides the asymptotic distribution of $D_{c}\n(r_1, r_2)$ under ${\cal H}_0$. Its proof is relegated to Section \ref{sec:proofs} in the Supplementary Material, where all the results of the paper are proved.

\begin{proposition} \label{Firstres}
	Fix $(r_1, r_2) \in \R^2$. Under ${\cal H}_0$, $ \sqrt{n} D_{c}\n(r_1, r_2)$ converges weakly as $\ny$ to a Gaussian random variable with mean zero and variance $V(r_1, r_2)\defin\E\big[\big\{\cos \big(r_1\vartheta^{(1)}+r_2\vartheta^{(2)}\big)-{\cal J}_{2c}(r_2)\cos \big(r_1\vartheta^{(1)}\big)-{\cal J}_{1c}(r_1)\cos \big(r_2\vartheta^{(2)}\big) +{\cal J}_{2s}(r_2)\sin \big(r_1\vartheta^{(1)}\big)+{\cal J}_{1s}(r_1)\sin \big(r_2\vartheta^{(2)}\big) \big\}^2\big]$.
\end{proposition}

The asymptotic normality of $\sqrt{n} D_{c}\n(r_1, r_2)$ does not depend on the distribution of the pair of random angles $(\vartheta^{(1)}, \vartheta^{(2)})$. A purely nonparametric test of independence can therefore be obtained on the basis of the Proposition \ref{Firstres}. Indeed, we can consider tests $\phi_{c}\n(r_1, r_2)$ rejecting the null hypothesis of independence at the asymptotic level $\alpha$ when
\begin{align}
	T_n(r_1, r_2)\defin\frac{n\big(D_{c}\n (r_1, r_2)\big)^2}{\hat{V}_n(r_1, r_2)} > \chi^2_{1; 1-\alpha},\label{eq:phiTn}
\end{align} 
where $\chi^2_{1; \nu}$ denotes the $\nu$th (lower) quantile of the chi-square distribution with $1$ degrees of freedom and $\hat{V}_n(r_1, r_2)$ is a consistent estimator of the variance term $V(r_1, r_2)$ defined in Proposition~\ref{Firstres}, such as its direct empirical version. Although being purely nonparametric, the tests $\phi_{c}\n(1, 1)$ and $\phi_{c}\n(1, -1)$ will enjoy certain local and asymptotic optimality properties. 

\subsection{Optimality and power against bivariate von Mises alternatives}

Consider the bivariate cosine von Mises model of \cite{Mardia2007}, characterized by densities of the form
\begin{align}
	\big(\vartheta^{(1)}, \vartheta^{(2)}\big) \mapsto C(\kappa_1, \kappa_2,\kappa_3) \exp\big\{\kappa_1 \cos \big(\vartheta^{(1)}\big)+ \kappa_2 \cos \big(\vartheta^{(2)}\big) +\kappa_3 \cos \big(\vartheta^{(1)}-\vartheta^{(2)}\big)\big\}, \label{cosvMF}
\end{align}
where $\kappa_1, \kappa_2 \geq 0$ are concentration parameters, $\kappa_3 \in \R$ is a parameter controlling the dependence, and $C(\kappa_1, \kappa_2,\kappa_3)$ is a normalizing constant. Note that, for the ease of our derivations, we flip the sign of $\kappa_3\in \R$ in \eqref{cosvMF} with respect to the original model parametrization. Following the terminology in \cite{Mardia2012}, density \eqref{cosvMF} is called the bivariate cosine model with \emph{positive interaction}. The same model with \emph{negative interaction} is obtained by replacing $\cos\big(\vartheta^{(1)}-\vartheta^{(2)}\big)$ with $\cos\big(\vartheta^{(1)}+\vartheta^{(2)}\big)$ in \eqref{cosvMF}. As stated in \cite{Mardia2007}, both models capture the correlations between the cosines and sines of the circular variables, though none is strictly associated with positive or negative correlations between angles. Indeed, the sign of `angular correlations' depends on $\kappa_3$, which affects asymmetrically the kind of dependence induced by \eqref{cosvMF}: positive values of $\kappa_3$ guarantee unimodality, with positive/negative angular correlation depending on the positive/negative interaction (Theorem 6.2 in \citealt{Mardia2012}; third column of Figure \ref{fig:2} in the Supplementary Material); negative $\kappa_3$ may generate bimodality distributed in an opposite correlation pattern to that of $\kappa_3>0$. Shifting of \eqref{cosvMF} can be achieved by replacing $\vartheta^{(j)}$ with $\vartheta^{(j)}-\mu^{(j)}$, for $\mu^{(j)}\in\mathbb{T}$, $j=1,2$. Location parameters do not affect the dependence form of \eqref{cosvMF}, yet they make it more cumbersome.\\

When $\kappa_3=0$, the marginals of \eqref{cosvMF} are independent centred von Mises distributions with concentrations $\kappa_1$ and $\kappa_2$, thus testing independence in this model reduces to testing ${\cal H}_0: \kappa_3=0$ against ${\cal H}_1: \kappa_3 \neq 0$. We show in Proposition \ref{optimality} that the tests $\phi_{c}\n(1, 1)$ and $\phi_{c}\n(1, -1)$ are locally and asymptotically maximin (see \citet[][Section 5]{Ley2017a} for a definition) for testing ${\cal H}_0: \kappa_3=0$ against ${\cal H}_1: \kappa_3 \neq 0$ within sequences of bivariate cosine models with negative and positive interaction, respectively. Recall that a test $\phi^*$ is called maximin in the class $\mathcal{C}_\alpha$ of level-$\alpha$ tests for some null hypothesis $\mathcal{H}_{0}$ against the alternative $\mathcal{H}_1$ if: (\textit{i}) $\phi^*$ has level $\alpha$; (\textit{ii}) the power of $\phi^*$ is such that
\begin{align*}
	\inf_{{\rm P} \in \mathcal{H}_1}\E_{{\rm P}}[\phi^*] \geq \sup_{\phi \in \mathcal{C}_\alpha} \inf_{{\rm P} \in \mathcal{H}_1} \E_{\rm P}[\phi].
\end{align*}

We denote by ${\rm P}_{(\kappa_1, \kappa_2, \kappa_3); -}\n$ and ${\rm P}_{(\kappa_1, \kappa_2, \kappa_3); +}\n$ the joint distributions of an independent and identically distributed sample $\big(\vartheta_1^{(1)}, \vartheta_1^{(2)}\big), \ldots, \big(\vartheta_n^{(1)}, \vartheta_n^{(2)}\big)$ from distribution \eqref{cosvMF}, respectively with negative and positive interaction. Obviously, ${\rm P}_{(\kappa_1, \kappa_2, 0); -}\n={\rm P}_{(\kappa_1, \kappa_2, 0); +}\n$, which is simply denoted as ${\rm P}_{(\kappa_1, \kappa_2, 0)}\n$.

\begin{proposition} \label{optimality}
	Letting $\tau_n$ be a bounded real sequence, the test $\phi_{c}\n(1, 1)$ is locally and asymptotically maximin for testing ${\cal H}_0: \cup_{\kappa_1 \geq 0} \cup_{\kappa_2 \geq 0} {\rm P}_{(\kappa_1, \kappa_2, 0)}\n$ against ${\cal H}_1:\cup_{\kappa_1 \geq 0} \cup_{\kappa_2 \geq 0} {\rm P}_{(\kappa_1, \kappa_2, n^{-1/2} \tau_n);-}\n$, while the test $\phi_{c}\n(1, -1)$ is locally and asymptotically maximin for testing ${\cal H}_0: \cup_{\kappa_1 \geq 0} \cup_{\kappa_2 \geq 0} {\rm P}_{(\kappa_1, \kappa_2, 0)}\n$ against ${\cal H}_1:\cup_{\kappa_1 \geq 0} \cup_{\kappa_2 \geq 0} {\rm P}_{(\kappa_1, \kappa_2, n^{-1/2} \tau_n);+}\n$. 
\end{proposition}

The nonparametric tests $\phi_{c}\n(1, 1)$ and $\phi_{c}\n(1, -1)$ therefore enjoy some parametric optimality properties for testing ${\cal H}_0: \kappa_3=0$ against ${\cal H}_1: \kappa_3 \neq 0$. Although the tests $\phi_{c}\n(r_1, r_2)$, $(r_1, r_2)\in\mathbb{R}^2$ do not enjoying such local and asymptotic optimality, it is easy to show that they enjoy non-trivial power against the contiguous alternatives ${\rm P}_{(\kappa_1, \kappa_2, n^{-1/2} \tau_n);+}\n$ and ${\rm P}_{(\kappa_1, \kappa_2, n^{-1/2} \tau_n);-}\n$, and can therefore be considered as reasonable tests for such alternatives.\\

Hitherto, we have assumed the sample comes from a circularly-centred random vector. Otherwise, the test statistic $T_n(r_1, r_2)$ in \eqref{eq:phiTn} has to be computed from the centred data $\vartheta_i^{(j)}-\mu^{(j)}$, $i=1, \ldots,n$, $j=1,2$; Proposition \ref{Firstres} then holds replacing the $\vartheta_i^{(j)}$'s and the $\vartheta^{(j)}$'s by $\vartheta_i^{(j)}-\mu^{(j)}$ and $\vartheta^{(j)}-\mu^{(j)}$, respectively, $i=1, \ldots, n$, $j=1,2$. Moreover, the local and asymptotic optimality obtained in Proposition \ref{optimality} also holds in the unspecified location case. Of course, the location parameters $\mu^{(1)}$ and $\mu^{(2)}$ are rarely known in practice so that they have to be estimated. This can be done using the sample circular means
\begin{align*}
	\hat{\mu}^{(j)}\defin\mathrm{atan2}\left(\frac{1}{n}\sum_{i=1}^n\sin\big(\vartheta_i^{(j)}\big),\frac{1}{n}\sum_{i=1}^n\cos\big(\vartheta_i^{(j)}\big)\right),\quad j=1,2.
\end{align*}
This estimation produces the centred sample
\begin{align}
	\big(\vartheta_1^{(1)}-\hat{\mu}^{(1)},\vartheta_1^{(2)}-\hat{\mu}^{(2)}\big), \ldots, \big(\vartheta_n^{(1)}-\hat{\mu}^{(1)},\vartheta_n^{(2)}-\hat{\mu}^{(2)}\big).\label{eq:centred}
\end{align}
When computed from this centred sample, the test statistic $T_n(r_1, r_2)$ in \eqref{eq:phiTn} is rotation invariant, which is a highly desirable property in the present toroidal context. We moreover have the following result.

\begin{proposition}\label{invariance}
	Denote by $\hat{D}_{c}\n(r_1, r_2)$ and ${D}_{c}\n(r_1, r_2)$ the quantities defined in \eqref{realpart}, but computed from the samples \eqref{eq:centred} and
	\begin{align*}
		\big(\vartheta_1^{(1)}-\mu^{(1)},\vartheta_1^{(2)}-\mu^{(2)}\big), \ldots, \big(\vartheta_n^{(1)}-\mu^{(1)},\vartheta_n^{(2)}-\mu^{(2)}\big),
	\end{align*}
	respectively. Then, provided that $\sqrt{n}\big(\hat{\mu}^{(j)}-\mu^{(j)}\big)=O_{\rm P}(1)$ as $\ny$, $j=1,2$, $\sqrt{n}(\hat{D}_{c}\n(r_1, r_2) -D_{c}\n(r_1, r_2))$ is $o_{\rm P}(1)$ as $\ny$.
\end{proposition}

Classical arguments similarly show that, provided that the data generating process is such that $\sqrt{n}\big(\hat{\mu}^{(j)}-\mu^{(j)}\big)=O_{\rm P}(1)$, $j=1,2$, the centring has no asymptotic effect on $\hat V_n(r_1,r_2)$ in \eqref{eq:phiTn}. Consequently, the centring step does not affect the asymptotic null distribution of $T_n(r_1, r_2)$ in \eqref{eq:phiTn}. Note that the same holds under contiguous alternatives. Since the centring of the sample is innocuous in terms of the asymptotic behaviour of \eqref{eq:phiTn} and it makes the test rotation invariant, this centring is implicitly assumed henceforth when applying the $\phi_c^{(n)}(r_1,r_2)$ test.\\

We conclude the section by pointing out that, while being of a nonparametric nature, the tests $\phi_{c}\n(r_1, r_2)$ are clearly designed to detect certain types of dependence (and not any kind of dependence): as seen in Proposition \ref{optimality}, the tests $\phi_{c}\n(1, \pm1)$ are particularly well-adapted to bivariate cosine von Mises alternatives that feature reflective symmetric marginal distributions. Note that working along the same lines, one could consider tests based on the sine empirical moments and show that some of their versions are locally and asymptotically optimal within specific parametric models. Rather than moving in this direction, in the following section we proceed towards tests of independence that are able to detect \emph{arbitrary} types of dependence.

\section{Omnibus tests}
\label{sec:extensions}

The well-known factorization property of characteristic functions entails that the null hypothesis of independence may equivalently be stated as 
\begin{align}\label{null1}
	\varphi(r_1,r_2)=\varphi_{1}(r_1) \varphi_{2}(r_2), \quad\text{for all } (r_1,r_2) \in \Z^2,
\end{align}
where $\varphi(r_1,r_2)\defin\E\big[e^{\mathrm{i}(r_1\vartheta^{(1)}+r_2\vartheta^{(2)})}\big]$, $\mathrm{i}\defin\sqrt{-1}$, is the joint characteristic function and $\varphi_{j}(r_j)\defin\E\big[e^{\mathrm{i} r_j \vartheta^{(j)}}\big]$ stands for the marginal characteristic function of $\vartheta^{(j)}$, $j=1,2$. Recall that, for random variables on the real line, \eqref{null1} needs to be considered for all $(r_1,r_2) \in \R^2$ while, due to periodicity, in the case of circular random variables, it is sufficient to consider the characteristic functions only for integer arguments. This is because the joint distribution of $\big(\vartheta^{(1)}, \vartheta^{(2)}\big)$ is identical to that of $\big(\vartheta^{(1)}+2\pi, \vartheta^{(2)}\big)$ and thus we have $\varphi(r_1,r_2)=e^{\mathrm{i} 2\pi r_1} \varphi(r_1,r_2)$, hence $r_1$ must be an integer, and likewise for $r_2$ \cite[Section 2.1]{Jammalamadaka2001}.\\

Based on $\big(\vartheta_1^{(1)}, \vartheta_1^{(2)}\big), \ldots, \big(\vartheta_n^{(1)}, \vartheta_n^{(2)}\big)$, the classical estimator of the joint characteristic function is
\begin{align}\label{ECF}
	\hat \varphi(r_1,r_2)\defin\frac{1}{n}\sum_{i=1}^{n} e^{ \mathrm{i}(r_1 \vartheta^{(1)}_{i}+r_2 \vartheta^{(2)}_{i})}, 
\end{align}
while the corresponding empirical marginals, say $\hat \varphi_{1}$ (respectively, $\hat \varphi_{2}$), can be obtained by setting $r_2=0$ ($r_1=0$) in \eqref{ECF}. Then, in view of \eqref{null1}, it is natural to consider the test statistics
\begin{align}\label{Dn}
	D\n(r_1,r_2)\defin \hat \varphi(r_1,r_2)-\hat \varphi_{1}(r_1)\hat \varphi_{2}(r_2),\quad (r_1,r_2)\in\mathbb{Z}^2,
\end{align}
as diagnostic components for independence. Notice that the quantity $D_{c}\n(r_1, r_2)$ defined in \eqref{realpart} is just the real part of $D\n(r_1,r_2)$, and consequently an extension of the tests studied in Section \ref{sec:nonparamtests} may be obtained by considering both the real and imaginary parts of $D\n(r_1,r_2)$ for multiple arguments $(r_1,r_2) \in \Z^2$. To this end, we define the vector 
\begin{align*}
	{\boldsymbol \Delta}_n\big(\boldsymbol{r}^{(c)},\boldsymbol{r}^{(s)}\big)\defin
	\left(D_c\n \big(r_{11}^{(c)},r_{12}^{(c)}\big),\ldots, D_c\n\big(r_{J1}^{(c)},r_{J2}^{(c)}\big),D_s\n\big(r_{11}^{(s)},r_{12}^{(s)}\big),\ldots,D_s\n\big(r_{K1}^{(s)},r_{K2}^{(s)}\big)\right)', 
\end{align*}
where $D_{c}\n(r_{1}, r_{2})$ and $D_{s}\n(r_{1}, r_{2})$ stand for the real and imaginary parts, respectively, of $D\n(r_{1}, r_{2})$. Using similar arguments as those in Section \ref{sec:nonparamtests}, it may be shown that $\sqrt{n} {\boldsymbol \Delta}_n\big(\boldsymbol{r}^{(c)},\boldsymbol{r}^{(s)}\big)$ is asymptotically a zero-mean multivariate Gaussian with some covariance matrix $\Sigb$ that is easily computable; see Section \ref{sec:Sigma} in the Supplementary Material. As a result, letting $\hat{\Sigb}$ be an invertible and consistent estimator of $\Sigb$, a very natural test $\phi\n\big(\boldsymbol{r}^{(c)},\boldsymbol{r}^{(s)}\big)$ rejects ${\cal H}_0$ for large values of $n \big({\boldsymbol \Delta}_n\big(\boldsymbol{r}^{(c)},\boldsymbol{r}^{(s)}\big)\big)' \hat{\Sigb}^{-1} {\boldsymbol \Delta}_n\big(\boldsymbol{r}^{(c)},\boldsymbol{r}^{(s)}\big)$. Note that some choices of $\boldsymbol{r}^{(c)}=\big(r_{11}^{(c)}, r_{12}^{(c)}, \ldots, r_{J1}^{(c)}, r_{J2}^{(c)}\big)' \in \Z^{2J}$ and $\boldsymbol{r}^{(s)}=\big(r_{11}^{(s)}, r_{12}^{(s)}, \ldots, r_{K1}^{(s)}, r_{K2}^{(s)}\big)' \in \Z^{2K}$ yield matrices $\Sigb$ that are invertible, some not. Note also that the particular case obtained by putting $J=2$ with $\big(r_{11}^{(c)}, r_{12}^{(c)}, r_{21}^{(c)}, r_{22}^{(c)}\big)=(1,-1,1,1)$ and $K=0$ (so that there is no `sine part' in ${\boldsymbol \Delta}_n\big(\boldsymbol{r}^{(c)},\boldsymbol{r}^{(s)}\big)$) yields a test that combines the two test statistics that are locally and asymptotically optimal against contiguous cosine von Mises alternatives with positive and negative dependence. An implicit centring of the sample is also assumed when applying $\phi\n\big(\boldsymbol{r}^{(c)},\boldsymbol{r}^{(s)}\big)$ as, analogously to the $\phi_c^{(n)}(r_1,r_2)$ test, this centring step is innocuous in terms of the asymptotic behaviour of the test and makes it rotation invariant.\\

While the tests $\phi\n\big(\boldsymbol{r}^{(c)},\boldsymbol{r}^{(s)}\big)$, with $\boldsymbol{r}^{(c)} \in \Z^{2J}$ and $\boldsymbol{r}^{(s)}\in \Z^{2K}$, are expected to have good power properties beyond the class of von Mises distributions for which $\phi\n_c(1,\pm1)$ is locally and asymptotically maximin, these tests are not `omnibus', i.e., they may potentially have trivial power against certain alternatives. In order to have an omnibus test, the uniqueness property of characteristic functions dictates that we must take into account all possible pairs $(r_1, r_2)\in \Z^2$. Consequently, we define a test criterion that rejects ${\cal H}_0$ for large values of 
\begin{align}\label{TS}
	T_{n,w}\defin n\sum_{r_1=-\infty}^\infty \sum_{r_2=-\infty}^\infty \left |D\n(r_1,r_2)\right|^2 w(r_1,r_2),
\end{align}
where $|\cdot|$ denotes the modulus of a complex number and $w:\Z^2 \rightarrow [0,\infty)$ is a weight function specified below. The following proposition formalizes the limit behaviour of $T_{n,w}$ against arbitrary deviations from the null hypothesis of independence.

\begin{proposition} 
	\label{Propext}
	Assume that $w$ in \eqref{TS} satisfies $\sum_{r_1=-\infty}^\infty \sum_{r_2=-\infty}^\infty w(r_1,r_2)<\infty$. Then,
	\begin{align} \label{ae}
		\frac{T_{n,w}}{n} \rightarrow {\cal{T}}_{w}\defin\sum_{r_1=-\infty}^\infty \sum_{r_2=-\infty}^\infty \left |\varphi(r_1,r_2)-\varphi_1(r_1)\varphi_2(r_2)\right|^2 w(r_1,r_2)
	\end{align}
	almost surely as $n \to \infty$. Moreover, ${\cal{T}}_{w}$ is strictly positive unless ${\cal H}_0$ holds true, a fact which entails strong consistency of the test that rejects ${\cal H}_0$ for large values of $T_{n,w}$. 
\end{proposition}

While $L_2$-type test statistics such as $T_{n,w}$ are omnibus, they typically have highly non-trivial asymptotic null distributions that essentially prevent their use as test criteria. We refer to \cite{Puri1977}, \cite{Shieh1994}, and \cite{Watson1967a} for analogous results; see also \citet[Section  8.9]{Jammalamadaka2001}. Nevertheless, it is straightforward to implement a permutation version of a test based on $T_{n,w}$.\\

The application of the test statistic would be further advanced if $T_{n,w}$ could be computed analytically. To this end, consider a weight function decomposed as $w(r_1,r_2)=v(r_1)v(r_2)$, with $v$ being a symmetric function about zero. Then, \eqref{TS} may be rewritten as (see Section \ref{sec:proofs} in the Supplementary Material)
\begin{align}
	T_{n,w}=&\;\frac{1}{n} \sum_{j,k=1}^{n}{\cal{J}}^{(v)}_c\big(\vartheta^{(1)}_{jk}\big){\cal{J}}^{(v)}_c\big(\vartheta^{(2)}_{jk}\big)+\frac{1}{n^3} \bigg[\sum_{j,k=1}^{n}{\cal{J}}^{(v)}_c\big(\vartheta^{(1)}_{jk}\big)\bigg] \bigg[\sum_{j,k=1}^{n}{\cal{J}}^{(v)}_c\big(\vartheta^{(2)}_{jk}\big)\bigg]\nonumber\\ &-\frac{2}{n^2} \sum_{j,k,\ell=1}^{n}{\cal{J}}^{(v)}_c\big(\vartheta^{(1)}_{jk}\big){\cal{J}}^{(v)}_c\big(\vartheta^{(2)}_{j\ell}\big),\label{TS2}
\end{align}where
\begin{align}\label{cv}
	{\cal{J}}^{(v)}_{c}(\vartheta)\defin\sum_{r=-\infty}^\infty \cos(r\vartheta)v(r),\\[-0.75cm]\nonumber
\end{align}
with $\vartheta^{(m)}_{jk}\defin\vartheta^{(m)}_{j}- \vartheta^{(m)}_{k}$, $j,k=1,\ldots,n$, $m=1,2$. Since $T_{n,w}$ only depends on the distances between observations, it is rotation-invariant without requiring a prior centring of the sample.\\

Moreover, if we consider any probability mass function on the non-negative integers and set $v$ equal to the symmetrized version of this function, then the series figuring in \eqref{cv} equals the real part of the characteristic function of that probability mass function, evaluated at $\vartheta$. A standard option is to choose the Poisson distribution, in which case
\begin{align}
	\label{poisson}
	{\cal{J}}^{(v)}_c(\vartheta)=\cos(\lambda \sin\vartheta)e^{\lambda(\cos\vartheta-1)},
\end{align}
where $\lambda$ is the Poisson parameter. Choosing $\lambda\in(0,\pi/2]$ guarantees the non-negativity of \eqref{poisson} for any $\vartheta\in\mathbb{T}$ (and also if $0<|\lambda|\leq \pi/2$). We denote by $T_{n,\lambda}$ the statistic \eqref{TS2} based on \eqref{poisson}. The test $\phi\n(\lambda)$ that rejects ${\cal H}_0$ for large values of $T_{n,\lambda}$ is implemented with a permutation approach that is described in Section \ref{sec:perm} of the Supplementary Material.

\section{Simulation study}
\label{sec:sim}

\subsection{Toroidal distributions considered}
\label{sec:toroidaldistributions}

To explore various shapes of dependence between $\vartheta^{(1)}$ and $\vartheta^{(2)}$, with a strength of dependence controlled by the value of a single parameter, we consider the four following joint parametric distributions of $\big(\vartheta^{(1)},\vartheta^{(2)}\big)$, all supported on $\mathbb{T}^2$:

\begin{enumerate}[label=(\textit{\roman*})]
	
	\item The ParaBolic distribution $\mathrm{PB}(p)$, defined by
	$\vartheta^{(1)} \sim \mathrm{Unif}\left(\mathbb{T}\right)$ and $\vartheta^{(2)} = 2 \big[p \big(\vartheta^{(1)}\big)^2 + (1 - p) U^2\big] / \pi - \pi$, where $U\sim \mathrm{Unif}\left(\mathbb{T}\right)$ is independent of $\vartheta^{(1)}$ and $p \in [0, 1]$. \label{mod1} 
	
	\item The (centred) Bivariate Wrapped Cauchy distribution as given in \cite{Pewsey2016}, denoted $\mathrm{BWC}(\rho_1, \rho_2, \rho)$ and with density being
	\begin{align*}
		\big(\vartheta^{(1)}, \vartheta^{(2)}\big) \mapsto c_0\big\{&c_1-c_2 \cos\big(\vartheta^{(1)}\big)-c_3 \cos\big(\vartheta^{(2)}\big) \nonumber\\
		&-c_4 \cos\big(\vartheta^{(1)}\big) \cos\big(\vartheta^{(2)}\big) -c_5 \sin\big(\vartheta^{(1)}\big) \sin\big(\vartheta^{(2)}\big)\big\}^{-1},
	\end{align*}
	where $c_j$, $j=0,\ldots,5$, are closed-form constants depending on $\rho_1,\rho_2,|\rho|\in[0,1)$. \label{mod2} 
	
	\item The (centred) Bivariate Cosine von Mises model with \emph{positive} interaction, denoted $\mathrm{BCvM}(\kappa_1, \kappa_2, \kappa_3)$ and with density described in Equation~\eqref{cosvMF}. \label{mod3} 
	
	\item The (centred) Bivariate von Mises by \cite{Shieh2005}, denoted $\mathrm{BvM}(\kappa_1, \kappa_2, \mu_g, \kappa_g)$ and with density
	\begin{align*}
		\big(\vartheta^{(1)}, \vartheta^{(2)}\big) \mapsto f_1\big(\vartheta^{(1)}\big)f_2\big(\vartheta^{(2)}\big) f_g\big(2\pi\big\{F_1\big(\vartheta^{(1)}\big)-F_2\big(\vartheta^{(2)}\big)\big\}\big),
	\end{align*}
	where $f_j$ and $F_j$ are respectively the marginal density and distribution functions of a zero-mean von Mises with concentration $\kappa_j\geq0$, $j=1,2$, and the link density $f_g$ is that of a von Mises with circular mean $\mu_g\in\mathbb{T}$ and concentration $\kappa_g\geq0$. \label{mod4} 
	
\end{enumerate}
The last parameter in each one of the four distributions controls the degree of dependence, with $p=\rho=\kappa_3=\kappa_g=0$ producing independence between $\vartheta^{(1)}$ and $\vartheta^{(2)}$.\\

Sampling from \ref{mod1} is straightforward. For \ref{mod3}, we used the function \texttt{rvmcos} from the \texttt{BAMBI} (v.~2.3.0) package \citep{Chakraborty2019}. One can simulate from \ref{mod4} using Algorithm~A for von Mises marginals in \cite{Shieh2005}. R codes for sampling \ref{mod2} and \ref{mod4} make use of package \texttt{circular} (v.~0.4-93) \citep{Agostinelli2017} and were kindly provided by Arthur Pewsey. They are available from the authors. Figure~\ref{fig:2} in the Supplementary Material shows different scatterplots obtained from the considered distributions.

\subsection{Empirical powers}
\label{subsec:empiricalpowers}

We investigate the empirical size and power of our three families of tests. More specifically, we consider the tests based on statistics $T_n(\boldsymbol{r}_1)$ and $T_n(\boldsymbol{r}_2)$ with $\boldsymbol{r}_1=(1,1)$ and $\boldsymbol{r}_2=(1,-1)$, $\boldsymbol{\Delta}_n\equiv\boldsymbol{\Delta}_n(\boldsymbol{r}^{(c)},\boldsymbol{r}^{(s)})$ with $\boldsymbol{r}^{(c)}=(1,-1,1,1)$ and $K=0$, and $T_{n,\lambda}$ for $\lambda\in\{0.1, 0.5, 1.0,2.0\}$. We also consider three competitors, namely the test based on the weighted $U$-statistic of \citet[p.~737]{Shieh1994}, denoted by $U_n$, the correlation test of \citet[p.~1835]{Zhan2019} based on the statistic $\hat \rho_0$, and the omnibus test of \cite{Rothman1971} based on the integrated empirical independence process denoted by~$C_n$.\\

The empirical power of these tests is compared by generating $M=10^5$ independent samples of sizes $n=20$ and $n=50$ from the distributions \ref{mod1}--\ref{mod4}, for varying dependence strengths. Results for a significance level $\alpha=5\%$ are summarised in Table~\ref{tab:4.5.1} for $n=50$ below and in Table~\ref{tab:4.5.2} in the Supplementary Material for $n=20$. In these tables, the first row in each panel corresponds to the independence case, while subsequent rows represent increasing dependence strength. The extreme cases $p=1$ and $\rho=1$ give functional dependence. We proceed as follows to compute critical values under ${\cal H}_0$. For a given sample size $n$, and a given bivariate parametric alternative distribution $\mathcal{D}(\theta)$, we generate two independent samples $\big(\vartheta^{(1)}_1,\vartheta^{(2)}_1\big),\ldots,\big(\vartheta^{(1)}_n,\vartheta^{(2)}_n\big)$ and $\big(\tilde{\vartheta}^{(1)}_1,\tilde{\vartheta}^{(2)}_1\big),\ldots,\big(\tilde{\vartheta}^{(1)}_n,\tilde{\vartheta}^{(2)}_n\big)$ from $\mathcal{D}(\theta)$. Critical values are then obtained by computing empirical quantiles from the sample $\big(\vartheta^{(1)}_1,\tilde{\vartheta}^{(2)}_1\big),\ldots,\big(\vartheta^{(1)}_n,\tilde{\vartheta}^{(2)}_n\big)$. While this necessitates to generate two samples, it is much faster than relying on a permutation approach. Moreover, this ensures that our empirical power values measure an ability to detect dependence by completely disregarding any potential marginal effect since the marginal distributions of $\big(\vartheta^{(1)},\tilde{\vartheta}^{(2)}\big)$ are the same as those of  $\big(\vartheta^{(1)},\vartheta^{(2)}\big)$, under the null and the alternative, respectively. We present in Appendix~\ref{sec:moresimus} an extensive simulation study showing that this much faster approach is equivalent, in terms of comparing the power values of the ten tests under scrutiny, to obtaining by permutations the critical values. Both approaches lead to very close power values for all four scenarios considered.

\begin{table}[h]
	\centering
	\footnotesize
		\begin{tabular}{cc|rrrrrrrrrr}
			\toprule
			&  & $T_n(\boldsymbol{r}_1)$ & $T_n(\boldsymbol{r}_2)$ & $\boldsymbol{\Delta}_n$ & $T_{n,0.1}$ & $T_{n,0.5}$ & $T_{n,1.0}$ & $T_{n,2.0}$ & $U_n$ & $\hat{\rho}_0$ & $C_n$ \\
			\midrule
			\multirow{6}{*}{$p$} & 0.0 & 5.00 & 5.05 & 4.95 & 4.95 & 4.97 & 4.99 & 5.00 & 4.74 & 4.86 & 4.79 \\
			& 0.2 & 8.87 & 8.88 & 10.47 & 12.54 & 16.37 & 22.81 & \textbf{30.48} & 11.17 & 12.91 & 12.17 \\
			& 0.4 & 39.03 & 38.88 & 47.93 & 75.72 & 82.01 & \textbf{85.90} & 79.04 & 18.07 & 25.01 & 21.26 \\
			& 0.6 & 69.00 & 68.85 & 87.13 & \textbf{99.98} & \textbf{99.98} & \textbf{99.98} & 99.66 & 33.51 & 46.85 & 52.92 \\
			& 0.8 & 73.27 & 72.84 & 95.66 & \textbf{100.00} & \textbf{100.00} & \textbf{100.00} & \textbf{100.00} & 56.63 & 62.09 & 99.89 \\
			& 1.0 & 66.87 & 66.93 & 83.73 & \textbf{100.00} & \textbf{100.00} & \textbf{100.00} & \textbf{100.00} & 71.16 & 70.48 & \textbf{100.00} \\ 
			\midrule
			\multirow{5}{*}{$\rho$} & 0.0 & 5.04 & 4.85 & 4.94 & 4.96 & 5.01 & 5.03 & 5.06 & 5.03 & 5.01 & 4.93 \\
			& 0.2 & 30.15 & 4.74 & 23.26 & \textbf{30.82} & 30.03 & 26.60 & 14.84 & 18.63 & 20.68 & 3.53 \\
			& 0.4 & 74.85 & 4.96 & 69.17 & \textbf{91.26} & 90.70 & 87.91 & 66.01 & 77.59 & 80.21 & 28.50 \\
			& 0.6 & 91.11 & 5.21 & 89.59 & \textbf{99.98} & \textbf{99.98} & \textbf{99.97} & 99.40 & 99.71 & 99.77 & 95.59 \\
			& 0.8 & 97.55 & 5.14 & 97.17 & \textbf{100.00} & \textbf{100.00} & \textbf{100.00} & \textbf{100.00} & \textbf{100.00} & \textbf{100.00} & \textbf{100.00} \\
			\midrule
			\multirow{6}{*}{$\kappa_3$} & 0.0 & 4.86 & 4.93 & 4.91 & 4.98 & 4.99 & 5.11 & 5.04 & 5.09 & 4.88 & 4.98 \\
			& 0.5 & 10.07 & \textbf{39.46} & 27.43 & 32.74 & 30.37 & 24.92 & 13.22 & 13.20 & 30.75 & 20.10 \\
			& 1.0 & 21.70 & \textbf{88.71} & 77.86 & 81.41 & 78.24 & 69.57 & 40.01 & 41.25 & 78.11 & 57.48 \\
			& 1.5 & 35.22 & \textbf{99.11} & 96.89 & 97.35 & 96.42 & 93.08 & 71.47 & 72.01 & 96.65 & 84.89 \\
			& 2.0 & 45.13 & \textbf{99.94} & 99.62 & 99.61 & 99.45 & 98.77 & 90.13 & 88.74 & 99.63 & 95.55 \\
			& 3.0 & 56.22 & \textbf{100.00} & \textbf{100.00} & 99.99 & 99.98 & 99.96 & 99.23 & 98.47 & 99.99 & 99.66 \\
			\midrule
			\multirow{6}{*}{$\kappa_g$} & 0.0 & 4.79 & 4.87 & 4.79 & 4.94 & 4.82 & 4.84 & 4.93 & 4.94 & 4.95 & 5.05 \\
			& 0.5 & 6.49 & \textbf{43.85} & 31.75 & 34.98 & 40.05 & 42.11 & 32.66 & 27.76 & 43.16 & 41.95 \\
			& 1.0 & 8.84 & 93.57 & 88.53 & 89.49 & 94.65 & \textbf{96.08} & 90.66 & 88.02 & 94.06 & 95.20 \\
			& 1.5 & 10.85 & 99.74 & 99.41 & 99.68 & 99.96 & \textbf{99.98} & 99.86 & 99.71 & 99.90 & 99.96 \\
			& 2.0 & 12.77 & 99.97 & 99.96 & \textbf{100.00} & \textbf{100.00} & \textbf{100.00} & \textbf{100.00} & \textbf{100.00} & \textbf{100.00} & \textbf{100.00} \\
			& 3.0 & 15.27 & \textbf{100.00} & \textbf{100.00} & \textbf{100.00} & \textbf{100.00} & \textbf{100.00} & \textbf{100.00} & \textbf{100.00} & \textbf{100.00} & \textbf{100.00} \\
			\bottomrule
		\end{tabular}
	\caption{\small Empirical level and power (in \%) for the distributions $\mathrm{PB}(p)$, $\mathrm{BWC}(0.1,0.1,-\rho)$, $\mathrm{BCvM}(1,1,\kappa_3)$, and $\mathrm{BvM}(1, 1, 0, \kappa_g)$ (top to bottom), for $\alpha=5\%$ and $n=50$. On each row, the largest power value is in bold, and any other power value falling in the \cite{Wilson1927}'s $95\%$ binomial confidence interval for the theoretical power of this best test is also in bold.}\label{tab:4.5.1}
\end{table}

A value of empirical level outside the interval $[4.86,5.14]$ indicates that the nominal level ($5\%$) does not fall within the corresponding realized $95\%$ confidence interval. Given that $80$ empirical levels were computed, a Bonferroni correction permits to extend the acceptable range to be within $[4.76,5.24]$. The only observed marked discrepancy between nominal and empirical levels (i.e., $4.74$\%) occurs for the $U_n$-based test, the reason being that its statistic is a discrete random variable. \\

The following conclusions can be drawn from Tables~\ref{tab:4.5.1} and \ref{tab:4.5.2}:
\begin{enumerate}
	
	\item The optimality of $\phi\n_c(1,-1)$ is corroborated for alternatives \ref{mod3}. In general, $\phi\n_c(1,-1)$ has a reasonable power against positive-correlation alternatives \ref{mod3} and \ref{mod4}, while it has very low power against negative-correlation alternatives~\ref{mod2}. An opposite behaviour for $\phi\n_c(1,1)$ is evidenced.
	
	\item The $\phi\n((1,-1,1,1))$ test behaves as expected on merging the benefits of $\phi\n_c(1,-1)$ and $\phi\n_c(1,1)$, providing competitive powers (in particular, against the tests based on $U_n$, $\hat\rho_0$, and $C_n$) in all scenarios and against positive/negative correlation. It suffers a moderate loss of power with respect to the best-performing test among $\phi\n_c(1,-1)$ and $\phi\n_c(1,1)$.
	
	\item $\phi\n(\lambda)$ is a very competitive test overall. For at least one choice of $\lambda\in\{0.1,0.5,1.0\}$ per simulation scenario, it dominates the rest of the tests for distributions \ref{mod1}, \ref{mod2}, and \ref{mod4} or offers a competitive power for distribution \ref{mod3}. In particular, $\phi\n(\lambda)$ dominates the three competing tests based on $U_n$, $\hat\rho_0$, and $C_n$, for at least one choice of $\lambda\in\{0.1,0.5,1.0\}$ and for all scenarios. This dominance is more marked when comparing to the only competing omnibus test, the one based on $C_n$.
	
	\item The choice of $\lambda$ is influential on the power of $\phi\n(\lambda)$. The choice $\lambda=2$ is seen to be systematically worse, which might be explained by the fact that, in this case, the kernel~\eqref{poisson} can be negative. Therefore, the power of $\phi\n(\lambda)$ might be drained by reducing the value of $T_{n,\lambda}$ for certain pairwise angles $\vartheta_{jk}^{(\ell)}$, $j,k=1,\ldots,n$, $\ell=1,2$.
	
	\item The three competing tests based on $U_n$, $\hat\rho_0$, and $C_n$ have a comparative poorer performance in scenario \ref{mod1}, which does not have a positive/negative-dependence pattern. In this case, our three tests clearly outperform the competition by a large margin.
	
\end{enumerate}

Overall, we recommend the use of the test $\phi^{(n)}(\lambda)$ for $\lambda\in\{0.1,1.0\}$.

\section{Data applications}
\label{sec:data}

\subsection{Long-period comets}
\label{sec:comets}

Long-period comets are thought to originate in the Oort cloud, a widely accepted model posing the existence of a roughly spherical reservoir of icy planetesimals in the limits of the Solar System. It is believed that these icy planetesimals become long-period comets when randomly captured in heliocentric orbits due to the effect of several gravitational forces \citep[e.g., Section 5 and Section 7.2 in ][]{Dones2015}. This conjectured origin explains the highly-characteristic nearly-isotropic distribution of the long-period comets' orbits \citep[e.g.,][]{Wiegert1999}. Such distribution is markedly different from that of short-period comets, who originate at the flattened Kuiper belt and whose orbits cluster about the ecliptic plane.\\

An orbit with \textit{inclination} $i\in[0,\pi]$ and \textit{longitude of the ascending node} $\Omega\in[0,2\pi)$ has directed normal vector $(\sin(i)\sin(\Omega),-\sin(i)\cos(\Omega),\cos(i))'$ to the orbit's plane \citep[e.g.,][]{Jupp2003}. Using this parametrization, \cite{Garcia-Portugues2020b} concluded a rejection ($p$-values smaller than $0.0053$) of the uniformity of the orbits of long-period comets, based on the records of the JPL Small-Body Database Search Engine (\url{https://ssd.jpl.nasa.gov/sbdb_query.cgi}) as of May 2020. The rejection may be driven by a truly non-uniform population or, according to the analysis in \cite{Jupp2003}, by the existence of significant observational bias on the available records. As \cite{Jupp2003} explain, bias is induced by how comet search programs maximize success detection chances by preferentially exploring regions about the ecliptic plane, as those are where most asteroids and short-period comets cluster.\\

A possible manifestation of observational bias, both in long- and short-period comets, is in the appearance of serial dependence in the orbits of observed comets. To assess the existence of such serial dependence, we investigated the lag-$1$ autocorrelation of the time series of $\Omega$. We used the lagged samples $(\Omega_{i},\Omega_{i+1})$, $i=1,\ldots,n-1$, with $n=445$ for long-period comets and $n=842$ for short-period comets (see Figure \ref{fig:1}). The dataset, which is the one employed in \cite{Garcia-Portugues2020b}, is available through the \texttt{comets} object of the \texttt{sphunif} R package \citep{Garcia-Portugues2020c}, and is sorted through the JPL's database ID, which is assigned chronologically based on the discovery of new comets.\\

\begin{figure}[h]
	\centering
	\includegraphics[width=0.33\textwidth]{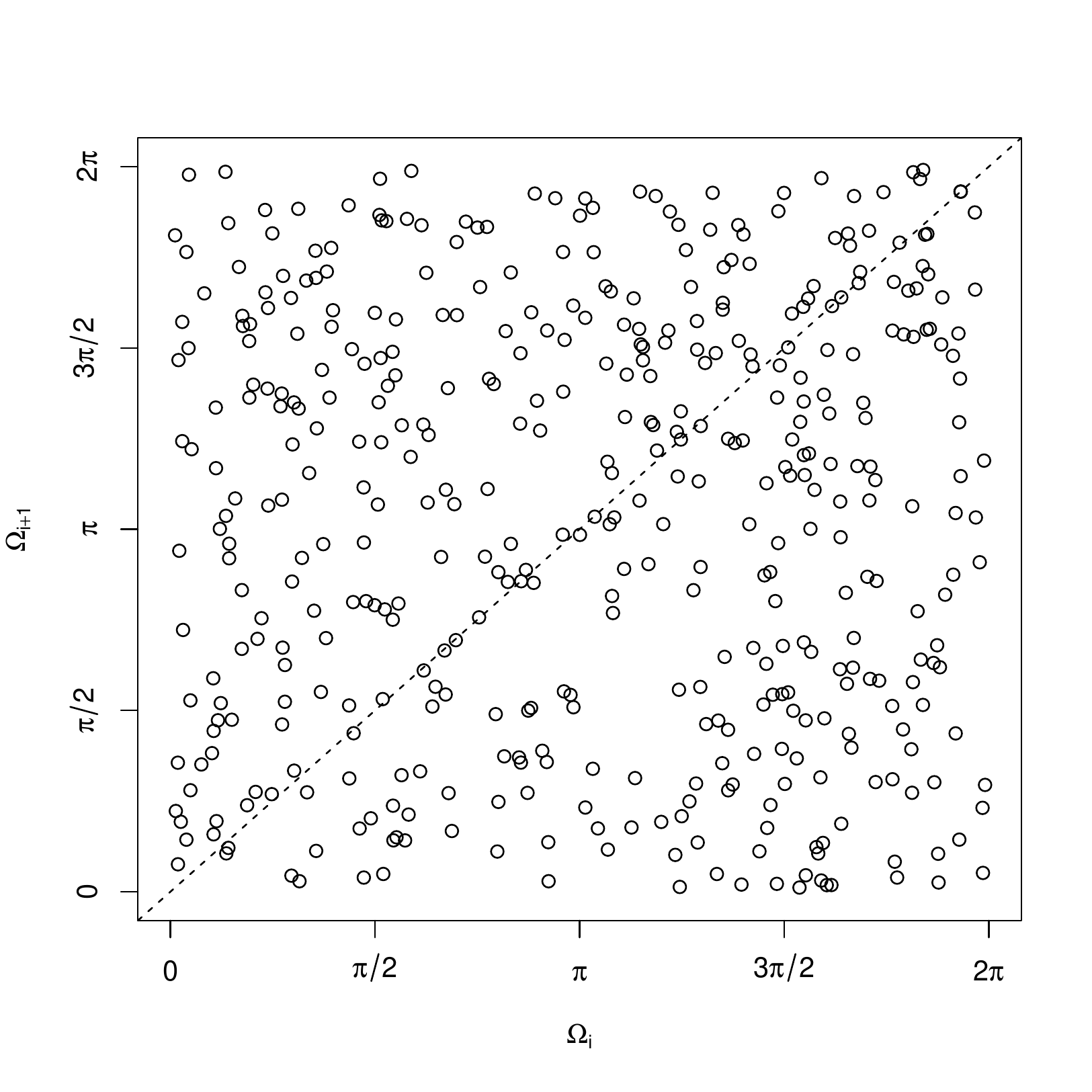}\includegraphics[width=0.33\textwidth]{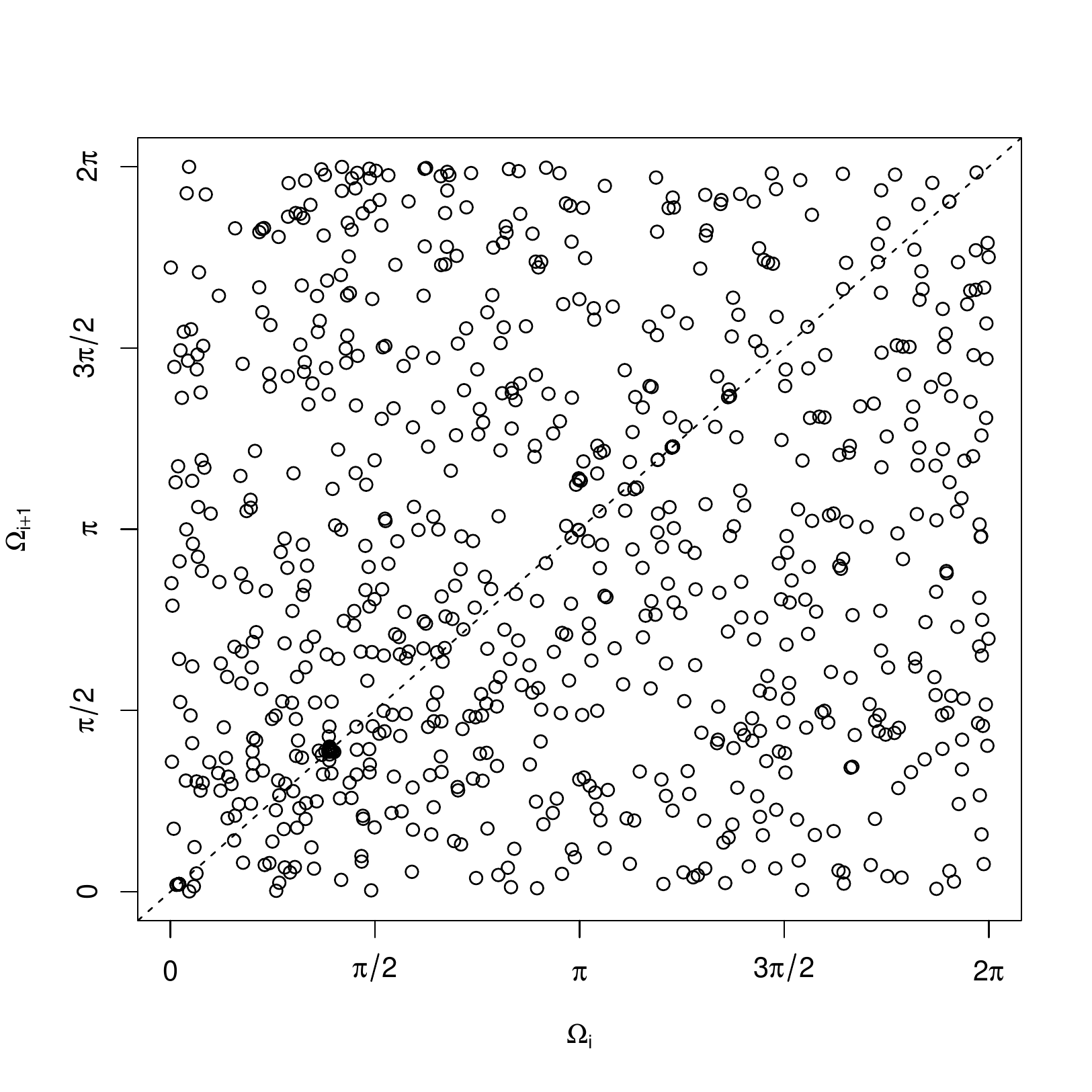}\includegraphics[width=0.33\textwidth]{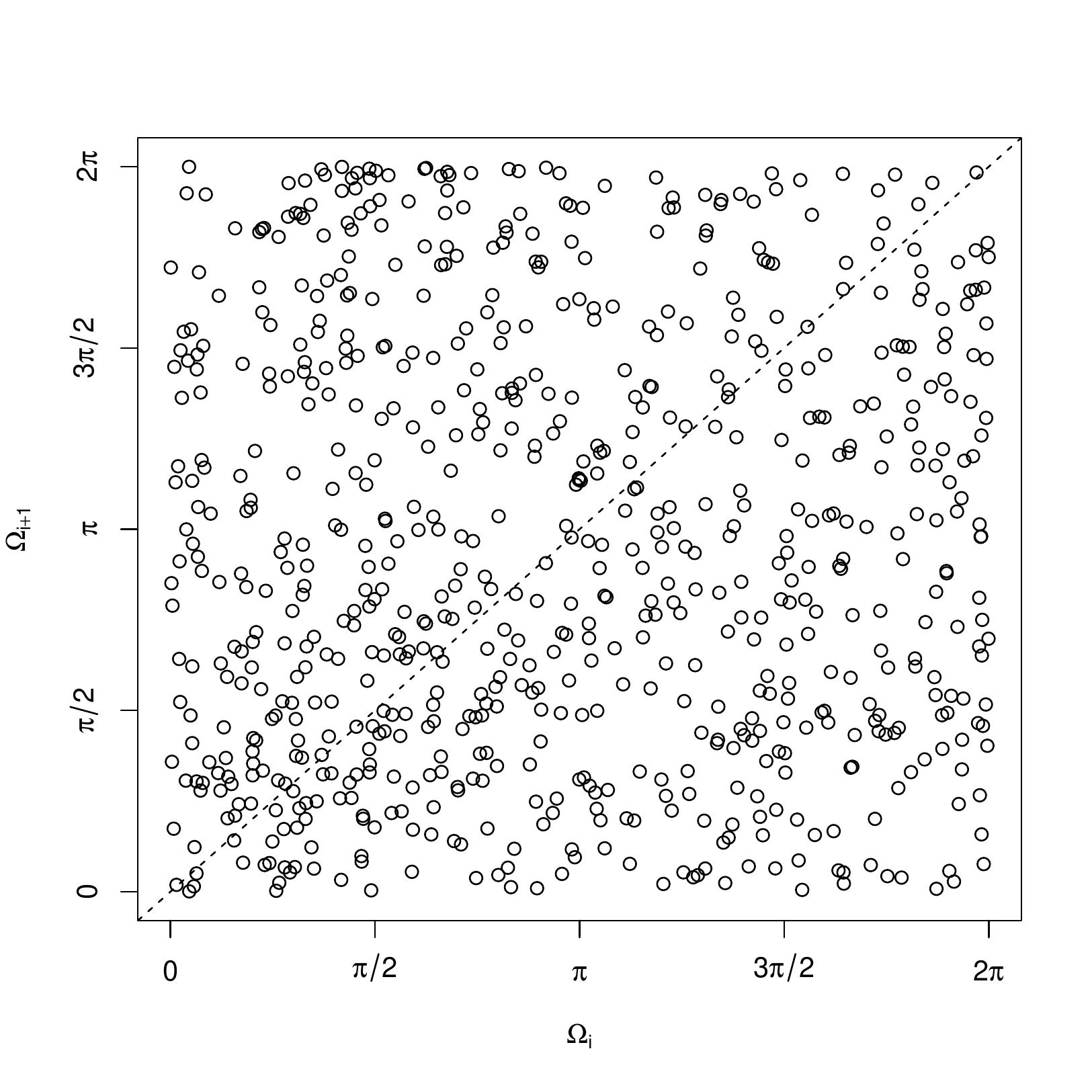}
	\caption{\small Scatterplots of $(\Omega_i,\Omega_{i+1})$ for long-period comets (left) and short-period comets (centre and right). The clusters appearing on the diagonal of the central plot disappear once the fragments of disintegrating comets are removed from the dataset (right plot). \label{fig:1}}
\end{figure}

The tests $\phi_{c}\n(1, 1)$, $\phi_{c}\n(1, -1)$, $\phi\n((1, -1,1, 1))$, $\phi\n(0.1)$, and $\phi\n(1)$ yielded $p$-values $0.6322$, $0.5334$, $0.9621$, $0.7795$, and $0.8849$ (using $10^4$ permutations for $\phi\n(\lambda)$), respectively, for the lagged sample of long-period comets. Therefore, no evidence against (negative or positive) autocorrelation is found, indicating that, if significant observational bias is present, it does not induce the most obvious forms of serial dependence on $\Omega$. For short-period comets, the $p$-values were $0.0001$, $3.7\times 10^{-8}$, $1.3\times 10^{-8}$, $0$, and $0$, thus signaling significant autocorrelation. A data inspection reveals that this rejection is a consequence of the clusters formed by fragments of disintegrating comets (see central plot of Figure \ref{fig:1}). For example, there is a sequence of $68$ records corresponding to fragments of the `73P/Schwassmann--Wachmann 3' comet. After removing $113$ fragment records, the tests gave $p$-values $0.1987$, $0.4170$, $0.6846$, $0.2249$, and $0.1850$, hence not rejecting the absence of autocorrelation on the longitudes of non-disintegrating short-period comets. The same test decisions were obtained when using lags of order two and three in the whole analysis.

\subsection{Wildfires}
\label{sec:wildfires}

\cite{Barros2012} identified the existence of preferential orientations of wildfires on $102$ characteristic watersheds of Portugal (see Figure \ref{fig:3}) determined in a data-driven fashion. Their analysis quantified annual wildfire orientations through the axial direction (e.g., North--South) of the first principal component of a wildfire perimeter. These perimeters were obtained from Landsat imagery of Portugal after the end of wildfire season and were then assigned to different watersheds according to the position of their centroids. Wildfire orientation is likely explained by dominant weather during the Portuguese wildfire season \citep{Barros2012} and is significantly associated with the size of burnt area \citep{Garcia-Portugues2014}.\\

We aim to formally address the existence of significant long-term and short-term temporal patterns in the Portuguese wildfire orientations. As in \cite{Garcia-Portugues2014}, we restrict to the $26,870$ wildfires mapped in 1985--2005 due to the higher resolution of satellite imagery for that period (minimum mapping unit of $5$ hectares). We then perform two data preprocessing steps. First, since a wildfire (axial) orientation is a $\pi$-periodic angular variable $\vartheta$ supported in $[0,\pi)$, we consider $2\vartheta$, a standard circular variable supported in $[0,2\pi)$. With this simple transformation, the angles $\{0, \pi/2, \pi, 3\pi/2\}$ represent the $\{\text{E--W}, \text{NE--SW}, \text{N--S}, \text{NW--SE}\}$ orientations, respectively. Second, we summarize the preferred orientation of the wildfires in each watershed by their weighted circular sample mean, with weights being the product between the proportion of explained variance and the burnt area of wildfire perimeter. The resulting dataset has $102$ representative wildfire orientations, shown in Figure \ref{fig:3} for 1986--1995 and 1996--2005.

\begin{figure}[h]
	\centering
	\includegraphics[height=0.35\textheight,clip,trim={3cm 3cm 7cm 2.75cm}]{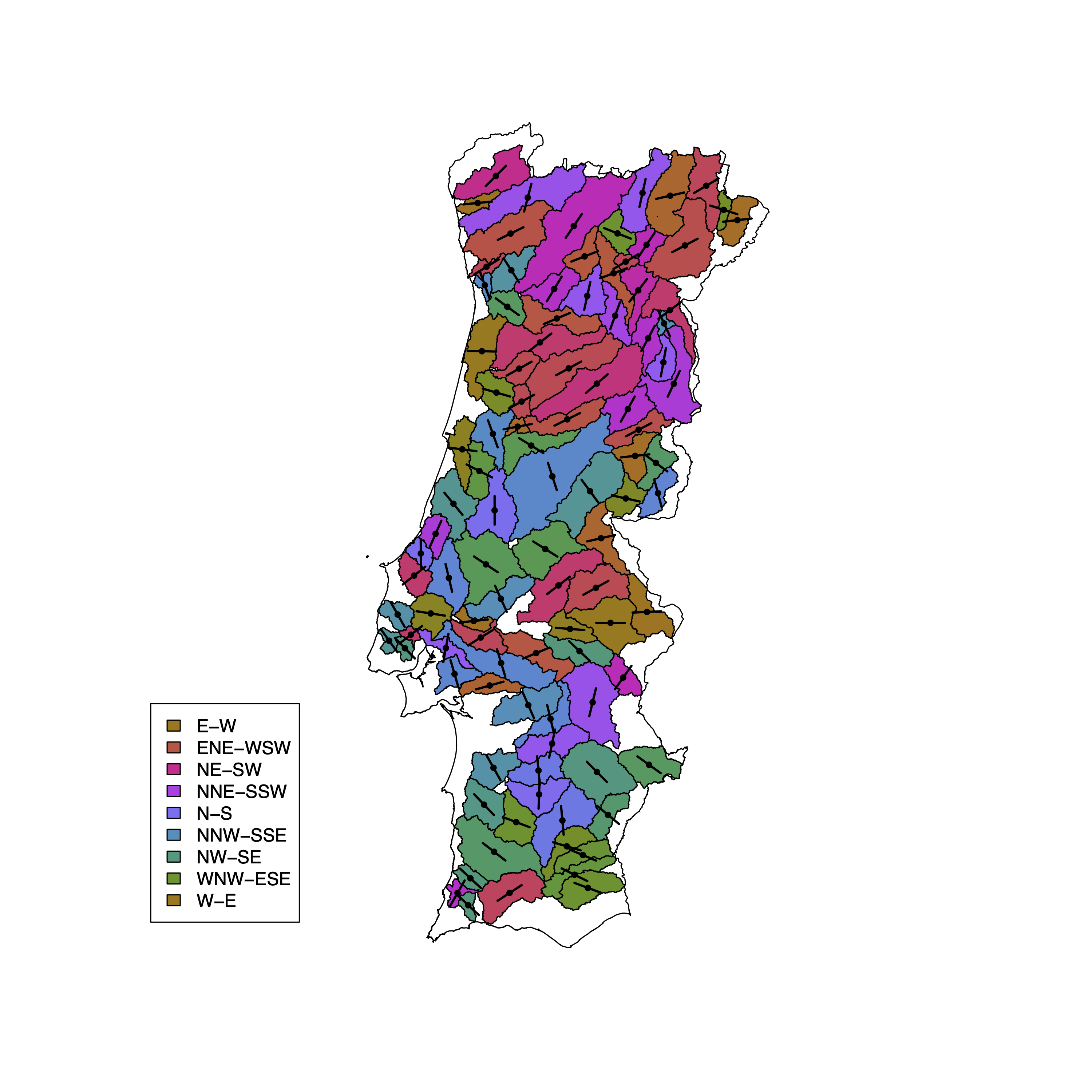}\includegraphics[height=0.35\textheight,clip,trim={7cm 3cm 7cm 2.75cm}]{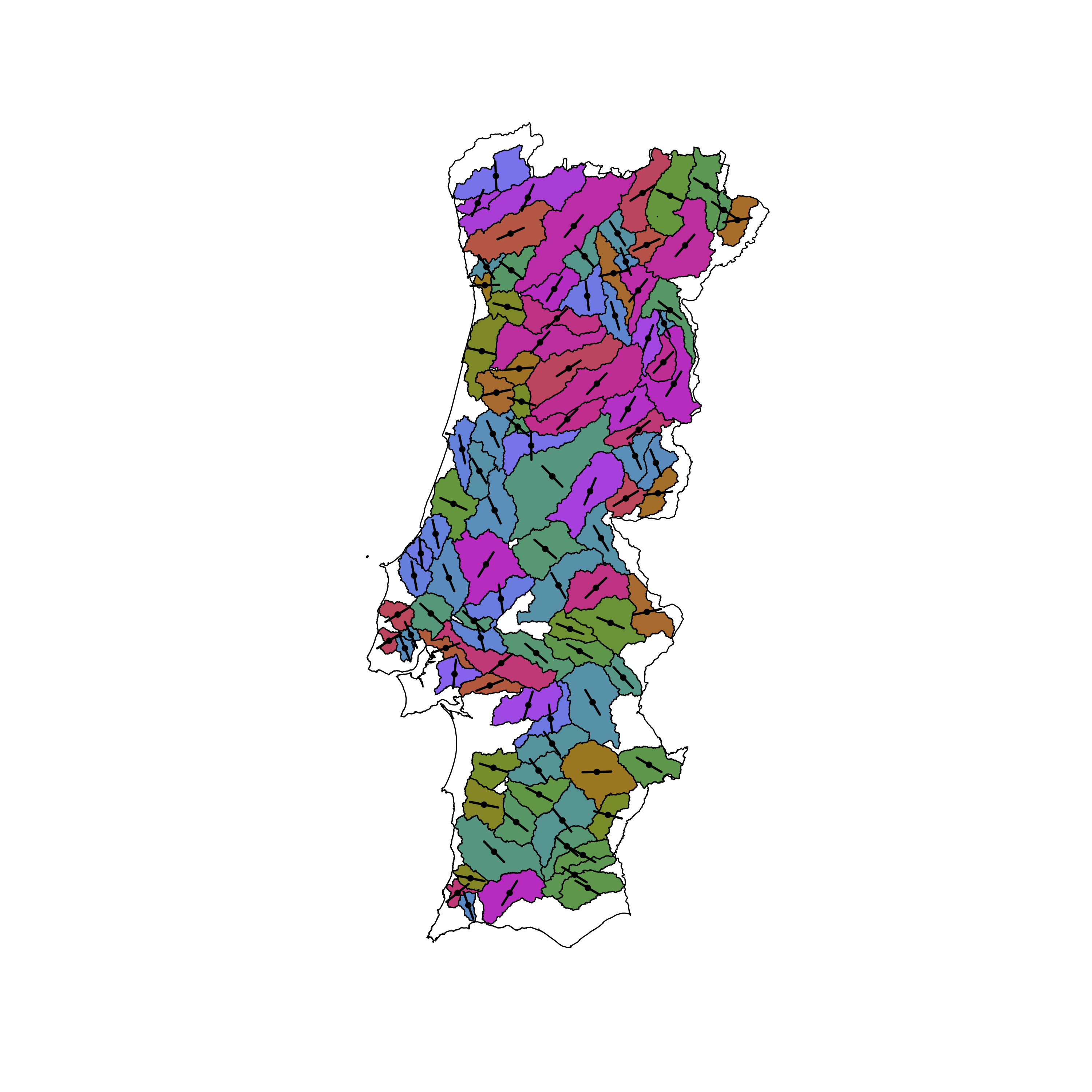}
	\caption{\small Weighted average orientations of the wildfires from 1986--1995 (left) and 1996--2005 (right), for each of the $102$ watersheds determined in \cite{Barros2012}. \label{fig:3}}
\end{figure}

When performed to the datasets displayed in Figure \ref{fig:3}, the tests $\phi_{c}\n(1, 1)$, $\phi_{c}\n(1, -1)$, $\phi\n((1, -1,1, 1))$, $\phi\n(0.1)$, and $\phi\n(1)$ yielded $p$-values $0.0593$, $0.0798$, $0.0730$, $0$, and $0.0003$. Therefore, significant long-term dependence is present in the orientation of wildfires. Short-term temporal dependence was also investigated by testing the null hypotheses of independence associated to the $20$ consecutive pairs of years in 1985--2005 and applying \cite{Benjamini2001}'s correction procedure. None of the (corrected) $p$-values of the five tests were below the $5\%$ significance level. For the $10\%$ significance level, only three $\phi_{c}\n(1, -1)$ tests and one $\phi\n(0.1)$ test were significant. To investigate mid-term temporal dependence, we repeated the analysis for pairs of consecutive periods of $5$-years ($12$ pairs) and $3$-years ($16$ pairs). The proportion of (corrected) $5\%$-significant $\phi_{c}\n(1, -1)$ tests raised to $0.5$ and $0.1875$, respectively, while again no $\phi_{c}\n(1, 1)$ tests were significant at any usual significance level. The corresponding proportions for the tests $\phi\n(0.1)$ and $\phi\n(1)$ were $0.75$ and $0.8333$ ($5$-years), and $0$ and $0.125$ ($3$-years). In conclusion, significant positive dependence of the orientations of wildfires is present among spans of $10$ and $5$ years, while no significant dependence is found on consecutive years. Both conclusions support the existence of drivers of the orientations in the long-term, such as dominant weather during the wildfire season \citep{Barros2012}.

\section*{Supplementary materials}

Supplementary materials provide the proofs of the stated results, describe the permutation algorithm for the $\phi\n(\lambda)$ test, and contain further simulation results.

\section*{Acknowledgements}

E. Garc\'ia-Portugu\'es acknowledges support by grants PGC2018-097284-B-100, IJCI-2017-32005, and MTM2016-76969-P from the Spain's Ministry of Economy and Competitiveness. All three grants were partially co-funded by the European Regional Development Fund. Part of this research was carried out while S. G. Meintanis was visiting P. Lafaye de Micheaux at UNSW, and hereby hospitality and financial support are sincerely acknowledged. T. Verdebout's research is supported by the Program of Concerted Research Actions (ARC) of the Universit\'{e} libre de Bruxelles. This research includes computations performed using the computational cluster Katana supported by Research Technology Services at UNSW Sydney.


\fi

\ifsupplement

\newpage
\title{Supplementary materials for ``Nonparametric tests of independence\newline for circular data based on trigonometric moments''}
\setlength{\droptitle}{-1cm}
\predate{}%
\postdate{}%
\date{}

\author{Eduardo Garc\'ia-Portugu\'es$^{1,8}$, Pierre Lafaye de Micheaux$^{2,3}$,\\
	Simos G. Meintanis$^{4,5}$, and Thomas Verdebout$^{6,7}$}
\footnotetext[1]{Department of Statistics, Carlos III University of Madrid (Spain).}
\footnotetext[2]{School of Mathematics and Statistics, University of New South Wales (Astralia).}
\footnotetext[3]{Institut Desbrest d'Epid\'{e}miologie et de Sant\'{e} Publique, Universit\'e Montpellier (France).}
\footnotetext[4]{Department of Economics, National and Kapodistrian University of Athens (Greece).}
\footnotetext[5]{Unit for Pure and Applied Analytics, North-West University (South Africa).}
\footnotetext[6]{D\'{e}partement de Math\'{e}matique, Universit\'{e} libre de Bruxelles (Belgium).}
\footnotetext[7]{ECARES, Universit\'{e} libre de Bruxelles (Belgium).}
\footnotetext[8]{Corresponding author. e-mail: \href{mailto:edgarcia@est-econ.uc3m.es}{edgarcia@est-econ.uc3m.es}.}
\maketitle

\begin{abstract}
	Supplementary materials provide the proofs of the stated results (Sections \ref{sec:proofs} and \ref{sec:Sigma}), describe the permutation algorithm for the $\phi\n(\lambda)$ test (Section \ref{sec:perm}), and contain further simulation results (Section \ref{sec:moresimus}).
\end{abstract}
\begin{flushleft}
	\small\textbf{Keywords:} Characteristic function; Circular data; Directional data; Independence; Trigonometric moments.
\end{flushleft}

\appendix

\section{Proofs}
\label{sec:proofs}

\begin{proof}{ of Proposition \ref{Firstres}}
	We readily have that
	\begin{align}\label{pr1step2}
		\sqrt{n}D_{c}\n(r_1, r_2) &= n^{-1/2} \sum_{i=1}^n \cos \big(r_1\vartheta_i^{(1)}+r_2\vartheta_i^{(2)}\big)- n^{1/2} \hat{\cal J}_{1c}(r_1) \hat{\cal J}_{2c}(r_2) + n^{1/2} \hat{\cal J}_{1s}(r_1)\hat{\cal J}_{2s}(r_2) \nonumber \\
		&= D_1\n - D_2\n+ D_3\n ,
	\end{align}
	where, in view of \eqref{JJ},
	\begin{align*}
		D_1\n&=n^{-1/2} \sum_{i=1}^n \big(\cos \big(r_1\vartheta_i^{(1)}+r_2\vartheta_i^{(2)}\big)- {\cal J}_{1c}(r_1){\cal J}_{2c}(r_2)+{\cal J}_{1s}(r_1){\cal J}_{2s}(r_2)\big),\\
		D_2\n&= n^{1/2} (\hat{\cal J}_{1c}(r_1) \hat{\cal J}_{2c}(r_2)-{\cal J}_{1c}(r_1){\cal J}_{2c}(r_2)),\\
		D_3\n&=n^{1/2} (\hat{\cal J}_{1s}(r_1)\hat{\cal J}_{2s}(r_2)-{\cal J}_{1s}(r_1){\cal J}_{2s}(r_2)).
	\end{align*}
	By the law of large numbers, $\hat{\cal J}(r)={\cal J}(r)+o_{\rm P}(1)$. Together with the Slutsky lemma, they entail that
	\begin{align*}
		D_2\n=&\; n^{1/2} (\hat{\cal J}_{1c}(r_1) \hat{\cal J}_{2c}(r_2)-{\cal J}_{1c}(r_1){\cal J}_{2c}(r_2)) \\
		=&\; \hat{\cal J}_{2c}(r_2) n^{1/2}(\hat{\cal J}_{1c}(r_1)-{\cal J}_{1c}(r_1))+ {\cal J}_{1c}(r_1) n^{1/2}(\hat{\cal J}_{2c}(r_2)-{\cal J}_{2c}(r_2)) \\
		=&\; {\cal J}_{2c}(r_2) n^{-1/2} \sum_{i=1}^n \big(\cos \big(r_1\vartheta_i^{(1)}\big)-{\cal J}_{1c}(r_1)\big)+ {\cal J}_{1c}(r_1) n^{-1/2} \sum_{i=1}^n \big(\cos \big(r_2\vartheta_i^{(2)}\big)-{\cal J}_{2c}(r_2)\big)\\
		&+o_{\rm P}(1)
	\end{align*}
	as $\ny$. Working along the same lines for $D_3\n$, we therefore obtain from \eqref{JJ} that
	\begin{align} 
		\sqrt{n} D_{c}\n(r_1, r_2)=&\; n^{-1/2} \sum_{i=1}^n \Big\{ \cos \big(r_1\vartheta_i^{(1)}+r_2\vartheta_i^{(2)}\big)- {\cal J}_{1c}(r_1){\cal J}_{2c}(r_2)+{\cal J}_{1s}(r_1){\cal J}_{2s}(r_2) \nonumber\\
		& -{\cal J}_{2c}(r_2) \big(\cos \big(r_1\vartheta_i^{(1)}\big)-{\cal J}_{1c}(r_1)\big)-{\cal J}_{1c}(r_1) \big(\cos \big(r_2\vartheta_i^{(2)}\big)-{\cal J}_{2c}(r_2)\big) \nonumber\\
		& +{\cal J}_{2s}(r_2) \big(\sin \big(r_1\vartheta_i^{(1)}\big)-{\cal J}_{1s}(r_1)\big)+{\cal J}_{1s}(r_1) \big(\sin \big(r_2\vartheta_i^{(2)}\big)-{\cal J}_{2s}(r_2)\big) \Big\}\nonumber\\
		&+ o_{\rm P}(1)\label{pr1step3}
	\end{align}
	as $\ny$. Under ${\cal H}_0$, the result then directly follows from \eqref{JJ} and the central limit theorem. Note that \eqref{pr1step3} is centred under ${\cal H}_0$, hence its asymptotic variance is simplified to
	\begin{align*}
		V(r_1,r_2)
		=&\;\E \big[\big\{\cos \big(r_1\vartheta^{(1)}+r_2\vartheta^{(2)}\big)- {\cal J}_{1c}(r_1){\cal J}_{2c}(r_2)+{\cal J}_{1s}(r_1){\cal J}_{2s}(r_2) \nonumber\\
		& -{\cal J}_{2c}(r_2) \big(\cos \big(r_1\vartheta^{(1)}\big)-{\cal J}_{1c}(r_1)\big)-{\cal J}_{1c}(r_1) \big(\cos \big(r_2\vartheta^{(2)}\big)-{\cal J}_{2c}(r_2)\big) \nonumber\\
		& +{\cal J}_{2s}(r_2) \big(\sin \big(r_1\vartheta^{(1)}\big)-{\cal J}_{1s}(r_1)\big)+{\cal J}_{1s}(r_1) \big(\sin \big(r_2\vartheta^{(2)}\big)-{\cal J}_{2s}(r_2)\big) \big\}^2\big]\\
		=&\;\E \big[\big\{\cos \big(r_1\vartheta^{(1)}+r_2\vartheta^{(2)}\big) -{\cal J}_{2c}(r_2) \cos \big(r_1\vartheta^{(1)}\big)-{\cal J}_{1c}(r_1) \cos \big(r_2\vartheta^{(2)}\big) \nonumber\\
		& +{\cal J}_{2s}(r_2) \big(\sin \big(r_1\vartheta^{(1)}\big)-{\cal J}_{1s}(r_1)\big)+{\cal J}_{1s}(r_1) \big(\sin \big(r_2\vartheta^{(2)}\big)-{\cal J}_{2s}(r_2)\big) \big\}^2\big].
	\end{align*}
\end{proof}

For the proof of Proposition \ref{optimality}, we consider local alternatives with positive interactions. The result for the local alternatives with negative interactions follows exactly along the same lines. The proof of Proposition \ref{optimality} relies on the local asymptotic normality property given in the following lemma, in which it is studied the asymptotic behaviour of likelihood ratio
\begin{align}\label{LR}
	\Lambda\n= \log \frac{d{\rm P}_{(\kappa_1+n^{-1/2} {k}_1\n, \kappa_2+n^{-1/2} {k}_2\n, n^{-1/2} {k}_3\n);+}\n}{d{\rm P}_{\kappa_1, \kappa_2, 0}\n}.
\end{align}

\begin{lemma} \label{LANres}
	Consider $\Lambda\n$ as defined in \eqref{LR}, where $\boldsymbol{k}_n=\big({k}_1\n, {k}_2\n, {k}_3\n\big)'$ is a bounded sequence in $\R^3$ such that $\kappa_1+n^{-1/2} {k}_1\n \geq 0$ and $\kappa_2+n^{-1/2} {k}_2\n \geq 0$. Then, under ${\rm P}_{\kappa_1, \kappa_2, 0}\n$,
	\begin{align*}
		\Lambda\n= {\boldsymbol{k}}_n' {\boldsymbol \Delta}\n- \frac{1}{2} {\boldsymbol k}_n' \Gamb{\boldsymbol k}_n+o_{\rm P}(1)
	\end{align*}
	as $\ny$, where ${\boldsymbol \Delta}\n=\big(\Delta_1\n, \Delta_2\n, \Delta_3\n\big)'$ with
	\begin{align*}
		{\Delta}_1\n&= n^{-1/2} \sum_{i=1}^n \big(\cos\big(\vartheta_i^{(1)}\big)-{\cal J}_{1c}(1)\big), \quad {\Delta}_2\n=n^{-1/2} \sum_{i=1}^n \big(\cos\big(\vartheta_i^{(2)}\big)-{\cal J}_{2c}(1)\big),\\
		{\Delta}_3\n&=n^{-1/2} \sum_{i=1}^n \big(\cos\big(\vartheta_i^{(1)}-\vartheta_i^{(2)}\big)- {\cal J}_{1c}(1){\cal J}_{2c}(1)\big),
	\end{align*}
	and
	\begin{align*}
		\Gamb= \left(\begin{array}{ccc} {\rm Var}\big[\cos\big(\vartheta_i^{(1)}\big)\big] &0 & {\cal J}_{2c}(1) {\rm Var}\big[\cos\big(\vartheta_i^{(1)}\big)\big] \\
			0 & {\rm Var}\big[\cos\big(\vartheta_i^{(2)}\big)\big] & {\cal J}_{1c}(1) {\rm Var}\big[\cos\big(\vartheta_i^{(2)}\big)\big] \\
			{\cal J}_{2c}(1) {\rm Var}\big[\cos\big(\vartheta_i^{(1)}\big)\big] & {\cal J}_{1c}(1) {\rm Var}\big[\cos\big(\vartheta_i^{(2)}\big)\big] & {\rm Var}\big[\cos\big(\vartheta_i^{(1)}-\vartheta_i^{(2)}\big)\big] \end{array} \right).
	\end{align*}
\end{lemma}

\begin{proof}{ of Lemma \ref{LANres}}
	The proof directly follows from the fact that the parametric model with density \eqref{cosvMF} is an exponential family with generating statistic ${\boldsymbol T}\big(\vartheta_i^{(1)},\vartheta_i^{(2)}\big)= \big(\cos\big(\vartheta_i^{(1)}\big), \cos\big(\vartheta_i^{(2)}\big), \cos\big(\vartheta_i^{(1)}-\vartheta_i^{(2)}\big)\big)'$ and is therefore $L_2$-differentiable \citep[e.g.,][Theorem 1.117 and Section 6]{Liese2008}.
\end{proof}

\begin{proof}{ of Proposition \ref{optimality}}
	Letting 
	\begin{align*}
		{\boldsymbol \Gamma}_{11}=&\;{\rm diag}\big({\rm Var}\big[\cos\big(\vartheta_i^{(1)})\big], {\rm Var}\big[\cos\big(\vartheta_i^{(2)}\big)\big]\big),\\
		{\boldsymbol \Gamma}_{12}=&\;\big({\cal J}_{2c}(1) {\rm Var}\big[\cos\big(\vartheta_i^{(1)}\big)\big], {\cal J}_{1c}(1) {\rm Var}\big[\cos\big(\vartheta_i^{(2)}\big)\big]\big)',
	\end{align*}
	it directly follows from Lemma \ref{LANres} that inference on $\kappa_3$ in the vicinity of $\kappa_3=0$ should be based on the \emph{efficient central sequence} {\citep[e.g.,][Section 5.2.4]{Ley2017a}}
	\begin{align}
		\Delta_{\rm eff}\n \defin&\; {\Delta}_3\n- {\boldsymbol \Gamma}_{12}' {\boldsymbol \Gamma}_{11}^{-1} \big({\Delta}_1\n, {\Delta}_2\n\big)' \nonumber \\
		=&\; n^{-1/2} \sum_{i=1}^n \big(\cos\big(\vartheta_i^{(1)}-\vartheta_i^{(2)}\big)- {\cal J}_{1c}(1){\cal J}_{2c}(1)\big) \nonumber \\ 
		& - {\cal J}_{2c}(1) n^{-1/2} \sum_{i=1}^n \big(\cos\big(\vartheta_i^{(1)}\big)-{\cal J}_{1c}(1)\big)-{\cal J}_{1c}(1) n^{-1/2} \sum_{i=1}^n \big(\cos\big(\vartheta_i^{(2)}\big)-{\cal J}_{2c}(1)\big). \nonumber
	\end{align}
	Since under ${\rm P}_{\kappa_1, \kappa_2, 0}\n$, ${\cal J}_{1s}(z)={\cal J}_{2s}(z)=0$ for all $z\in \Z$, it readily follows from ${\cal J}_{jc}(z)={\cal J}_{jc}(-z)$, $j=1,2$, and from \eqref{pr1step3} that $\sqrt{n} D_{c}\n(1, -1)= \Delta_{\rm eff}\n+o_{\rm P}(1)$ as $\ny$ under ${\rm P}_{\kappa_1, \kappa_2, 0}\n$. The result follows.
\end{proof}

\begin{proof}{ of Proposition \ref{invariance}}
	We write $\tilde{\cal J}_{jc}(r_j)\defin n^{-1} \sum_{i=1}^n \cos\big(r_j \big(\vartheta_i^{(j)}- \hat\mu^{(j)}\big)\big)$ and, as in the previous sections,
	\begin{align}
		\hat{\cal J}_{jc}(r_j)\defin n^{-1} \sum_{i=1}^n \cos\big(r_j \big(\vartheta_i^{(j)}- \mu^{(j)}\big)\big) \quad {\rm and} \quad {\cal J}_{jc}(r_j)\defin \E\big[\cos \big(r_j \big(\vartheta_i^{(j)}- \mu^{(j)}\big)\big)\big],\label{eq:center}
	\end{align}
	where $\hat\mu^{(1)}$ and $\hat\mu^{(2)}$ are root-$n$ consistent estimators of (well-identified) location parameters $\mu^{(1)}$ and $\mu^{(2)}$. We consider the asymptotic distribution of
	\begin{align*}
		\sqrt{n}\hat{D}_{c}\n(r_1, r_2) =&\; n^{-1/2} \sum_{i=1}^n \cos \big(r_1\big(\vartheta_i^{(1)}- \hat\mu^{(1)}\big)+r_2 \big(\vartheta_i^{(2)}- \hat\mu^{(2)}\big)\big)- n^{1/2} \tilde{\cal J}_{1c}(r_1) \tilde{\cal J}_{2c}(r_2) \\
		&+ n^{1/2} \tilde{\cal J}_{1s}(r_1)\tilde{\cal J}_{2s}(r_2) \\
		=&\; \hat{D}_1\n - \hat{D}_2\n+ \hat{D}_3\n,
	\end{align*}
	where
	\begin{align*} 
		\hat{D}_1\n&=n^{-1/2} \sum_{i=1}^n \cos \big(r_1\big(\vartheta_i^{(1)}- \hat\mu^{(1)}\big)+r_2 \big(\vartheta_i^{(2)}- \hat\mu^{(2)}\big)\big)- {\cal J}_{1c}(r_1){\cal J}_{2c}(r_2)+{\cal J}_{1s}(r_1){\cal J}_{2s}(r_2)), \\
		\hat{D}_2\n&= n^{1/2} (\tilde{\cal J}_{1c}(r_1) \tilde{\cal J}_{2c}(r_2)-{\cal J}_{1c}(r_1){\cal J}_{2c}(r_2)), \\
		\hat{D}_3\n&=n^{1/2} (\tilde{\cal J}_{1s}(r_1)\tilde{\cal J}_{2s}(r_2)-{\cal J}_{1s}(r_1){\cal J}_{2s}(r_2)).
	\end{align*}
	The combination of Lemmas \ref{Theelem} and \ref{Theelembis} below shows that 
	\begin{align*}
		\hat{D}_1\n - \hat{D}_2\n+ \hat{D}_3\n={D}_1\n - {D}_2\n+ {D}_3\n+o_{\rm P}(1)
	\end{align*}
	as $\ny$, where 
	\begin{align*} 
		{D}_1\n&=n^{-1/2} \sum_{i=1}^n \cos \big(r_1\big(\vartheta_i^{(1)}- \mu^{(1)}\big)+r_2 \big(\vartheta_i^{(2)}- \mu^{(2)}\big)\big)- {\cal J}_{1c}(r_1){\cal J}_{2c}(r_2)+{\cal J}_{1s}(r_1){\cal J}_{2s}(r_2)\big), \\
		{D}_2\n&= n^{1/2} (\hat{\cal J}_{1c}(r_1) \hat{\cal J}_{2c}(r_2)-{\cal J}_{1c}(r_1){\cal J}_{2c}(r_2)), \\
		{D}_3\n&=n^{1/2} (\hat{\cal J}_{1s}(r_1)\hat{\cal J}_{2s}(r_2)-{\cal J}_{1s}(r_1){\cal J}_{2s}(r_2)).
	\end{align*}
\end{proof}

\begin{lemma}\label{Theelem}
	Letting $\hat\mu^{(1)}$ and $\hat\mu^{(2)}$ be root-$n$ consistent estimators of location parameters $\mu^{(1)}$ and $\mu^{(2)}$, we have that, as $\ny$,
	\begin{align*}
		\hat{D}_2\n-{D}_2\n&= r_1 n^{1/2}\big(\hat\mu^{(1)}- \mu^{(1)}\big) {\cal J}_{1s}(r_1) {\cal J}_{2c}(r_2)+ r_2 n^{1/2}\big(\hat\mu^{(2)}- \mu^{(2)}\big) {\cal J}_{1c}(r_1) {\cal J}_{2s}(r_2)+ o_{\rm P}(1),\\
		\hat{D}_3\n-{D}_3\n&=- r_1 n^{1/2}\big(\hat\mu^{(1)}- \mu^{(1)}\big) {\cal J}_{1c}(r_1) {\cal J}_{2s}(r_2)- r_2 n^{1/2}\big(\hat\mu^{(2)}- \mu^{(2)}\big) {\cal J}_{1s}(r_1) {\cal J}_{2c}(r_2)+ o_{\rm P}(1),
	\end{align*}
	where the left hand side terms are defined in the proof of Proposition \ref{invariance}.
\end{lemma}

\begin{proof}{ of Lemma \ref{Theelem}}
	We use the notations $s_i^j\defin\sin (r_j \vartheta_i^{(j)})$, $c_i^j\defin\cos (r_j \vartheta_i^{(j)})$, $c^j\defin\cos(r_j \mu^{(j)})$, $\hat{c}^j\defin\cos(r_j \hat\mu^{(j)})$, $s^j\defin\sin(r_j \mu^{(j)})$, and $\hat{s}^j\defin\sin(r_j \hat\mu^{(j)})$.\\
	
	First note that 
	\begin{align}\label{diffthelem}
		\hat{D}_2\n-{D}_2\n &= n^{1/2}(\tilde{\cal J}_{1c}(r_1) \tilde{\cal J}_{2c}(r_2)-\hat{\cal J}_{1c}(r_1) \hat{\cal J}_{2c}(r_2)) \nonumber \\
		&= n^{1/2}(\tilde{\cal J}_{1c}(r_1)-\hat{\cal J}_{1c}(r_1)) \tilde{\cal J}_{2c}(r_2)+ n^{1/2}( \tilde{\cal J}_{2c}(r_2) -\hat{\cal J}_{2c}(r_2)) \hat{\cal J}_{1c}(r_1). 
	\end{align}
	Now the delta method directly entails that for $j=1,2,$ as $\ny$
	\begin{align} 
		n^{1/2}\big(\hat{c}^j-c^j\big)&=-r_j s^j n^{1/2}\big( \hat\mu^{(j)}- \mu^{(j)}\big)+o_{\rm P}(1),\label{deltacos}\\
		n^{1/2}\big(\hat{s}^j-s^j\big)&= r_j c^j n^{1/2}\big( \hat\mu^{(j)}- \mu^{(j)}\big)+o_{\rm P}(1).\label{deltasin}
	\end{align}
	Now, for $j=1,2$, classical trigonometric identities, the Slutsky lemma, and \eqref{deltacos} and \eqref{deltasin}, yield
	\begin{align*}
		n^{1/2}\big(\tilde{\cal J}_{jc}(r_j)-\hat{\cal J}_{jc}(r_j)\big) &=n^{-1/2} \sum_{i=1}^n \Big\{\cos\big(r_j \big(\vartheta_i^{(j)}- \hat\mu^{(j)}\big)\big) - \cos\big(r_j \big(\vartheta_i^{(j)}- \mu^{(j)}\big)\big)\Big\} \\
		&= n^{-1/2} \sum_{i=1}^n \Big\{c_i^j \hat{c}^j+s_i^j \hat{s}^j-c_i^j {c}^j-s_i^j {s}^j \Big\}\\
		&= n^{1/2}(\hat{c}^j-c^j) \Big\{n^{-1} \sum_{i=1}^n c_i^j\Big\}+n^{1/2}\big(\hat{s}^j-s^j\big) \Big\{n^{-1} \sum_{i=1}^n s_i^j\Big\} \\
		&= r_j n^{1/2}\big( \hat\mu^{(j)}- \mu^{(j)}\big) \Big\{n^{-1} \sum_{i=1}^n \big(s_i^j c^j-c_i^j s^j\big) \Big\}+o_{\rm P}(1) \\
		&= r_j n^{1/2}\big( \hat\mu^{(j)}- \mu^{(j)}\big) {\cal J}_{js}(r_j)+o_{\rm P}(1)
	\end{align*}
	as $\ny$. Therefore, it directly follows from the Slutsky lemma and \eqref{diffthelem} that
	\begin{align*}
		\hat{D}_2\n - {D}_2\n=r_1 n^{1/2}\big( \hat\mu^{(1)}- \mu^{(1)}\big) {\cal J}_{1s}(r_1) {\cal J}_{2c}(r_2)+ r_2 n^{1/2}\big( \hat\mu^{(2)}- \mu^{(2)}\big) {\cal J}_{1c}(r_1) {\cal J}_{2s}(r_2)+o_{\rm P}(1)
	\end{align*}
	as $\ny$. The result follows along the same lines for $\hat{D}_3\n - {D}_3\n$.
\end{proof}

\begin{lemma}\label{Theelembis}
	Letting $\hat\mu^{(1)}$ and $\hat\mu^{(2)}$ be root-$n$ consistent estimators of location parameters $\mu^{(1)}$ and $\mu^{(2)}$, we have that, as $\ny$,
	\begin{align*}
		\hat{D}_1\n-{D}_1\n = \big(r_1 \sqrt{n} \big(\hat\mu^{(1)}-\mu^{(1)}\big)+ r_2 \sqrt{n} \big(\hat\mu^{(2)}-\mu^{(2)}\big)\big) ({\cal J}_{1s}(r_1){\cal J}_{2c}(r_2)+{\cal J}_{1c}(r_1){\cal J}_{2s}(r_2)) + o_{\rm P}(1).
	\end{align*}
\end{lemma}

\begin{proof}{ of Lemma \ref{Theelembis}}
	First note that simple computations yield
	\begin{align} \label{donemin}
		\hat{D}_1\n&-{D}_1\n\nonumber\\
		=&\; n^{-1/2} \sum_{i=1}^n \Big\{\cos\big(r_1\big(\vartheta_i^{(1)}- \hat\mu^{(1)}\big)+r_2\big(\vartheta_i^{(2)}- \hat\mu^{(2)}\big) \big)-\cos\big(r_1\big(\vartheta_i^{(1)}- \mu^{(1)}\big)+r_2\big(\vartheta_i^{(2)}- \mu^{(2)}\big) \big) \Big\}\nonumber \\
		=&\;\sqrt{n}\big(\hat{c}^1\hat{c}^2-{c}^1 {c}^2\big) \Big\{n^{-1} \sum_{i=1}^n \big(c_i^1 c_i^2- s_i^1 s_i^2\big)\Big\}+ \sqrt{n}\big(\hat{c}^1\hat{s}^2-{c}^1 {s}^2\big) \Big\{n^{-1} \sum_{i=1}^n \big(c_i^1 s_i^2+s_i^1 c_i^2\big) \Big\}\nonumber\\
		&+\sqrt{n}\big(\hat{s}^1\hat{c}^2-{s}^1 {c}^2\big) \Big\{n^{-1} \sum_{i=1}^n \big(s_i^1 c_i^2+c_i^1s_i^2\big)\Big\}+\sqrt{n}\big(\hat{s}^1\hat{s}^2-{s}^1 {s}^2\big) \Big\{n^{-1} \sum_{i=1}^n \big(s_i^1 s_i^2- c_i^1 c_i^2\big)\Big\}.
	\end{align}
	Considering only the first term in \eqref{donemin}, we have using \eqref{deltacos}, \eqref{deltasin} and the Slutsky lemma that
	\begin{align*}
		\sqrt{n}\big(\hat{c}^1\hat{c}^2&-{c}^1 {c}^2\big) \Big\{n^{-1} \sum_{i=1}^n \big(c_i^1 c_i^2- s_i^1 s_i^2\big)\Big\} \\
		=&\; \big(\hat{c}^1 \sqrt{n}\big(\hat{c}^2-{c}^2\big)+ {c}^2 \sqrt{n} \big(\hat{c}^1-{c}^1\big)\big) \Big\{n^{-1} \sum_{i=1}^n \big(c_i^1 c_i^2- s_i^1 s_i^2\big)\Big\} \\
		=&\; \big(-r_2 \sqrt{n} \big(\hat\mu^{(2)}-\mu^{(2)}\big) s^2c^1-r_1 \sqrt{n} \big(\hat\mu^{(1)}-\mu^{(1)}\big) s^1c^2\big) \Big\{n^{-1} \sum_{i=1}^n \big(c_i^1 c_i^2- s_i^1 s_i^2\big)\Big\}+o_{\rm P}(1)
	\end{align*}
	as $\ny$. Working similarly for all the terms in \eqref{donemin}, from the law of large numbers and the Slutsky lemma, we obtain
	\begin{align*}
		\hat{D}_1\n-{D}_1\n =&\; \big(r_1 \sqrt{n} \big(\hat\mu^{(1)}-\mu^{(1)}\big)+r_2 \sqrt{n} \big(\hat\mu^{(2)}-\mu^{(2)}\big)\big) \Big\{n^{-1} \sum_{i=1}^n\big(-c_i^1c_i^2 s^1 c^2-c_i^1s_i^2 s^1 s^2 \\
		&+s_i^1c_i^2 c^1 c^2+s_i^1s_i^2 c^1 s^2+s_i^1s_i^2 s^1 c^2-s_i^1c_i^2 s^1 s^2 +c_i^1s_i^2 c^1 c^2 -c_i^1c_i^2 c^1 s^2\big) \Big\}+o_{\rm P}(1) \\
		=&\; \big(r_1 \sqrt{n} \big(\hat\mu^{(1)}-\mu^{(1)}\big)+r_2 \sqrt{n} \big(\hat\mu^{(2)}-\mu^{(2)}\big)\big) \E\big[\sin\big(r_1\big(\vartheta_i^{(1)}- \mu^{(1)}\big)+r_2\big(\vartheta_i^{(2)}- \mu^{(2)}\big) \big)\big]\\
		&+o_{\rm P}(1)\\
		=&\; \big(r_1 \sqrt{n} \big(\hat\mu^{(1)}-\mu^{(1)}\big)+ r_2 \sqrt{n} \big(\hat\mu^{(2)}-\mu^{(2)}\big)\big) ({\cal J}_{1s}(r_1){\cal J}_{2c}(r_2)+{\cal J}_{1c}(r_1){\cal J}_{2s}(r_2))\\
		&+ o_{\rm P}(1),
	\end{align*}
	which is the desired result.
\end{proof}

\begin{proof}{ of Proposition \ref{Propext}}
	From the strong law of large numbers, we have that, for each $(r_1,r_2) \in \mathbb Z^2$, $\hat \varphi(r_1,r_2) \rightarrow \varphi(r_1,r_2)$ almost surely as $n \to \infty$, and likewise for $\hat \varphi_m$, $m=1,2$. Therefore, \eqref{ae} follows from Lebesgue's dominated convergence theorem since $\left |D_n(r_1,r_2)\right|^2 \leq 4$. Moreover, in view of \eqref{null1}, the almost sure limit ${\cal{T}}_{w}$ in \eqref{ae} is positive unless the null hypothesis of independence holds true, which in turn implies that, under alternatives, $T_{n,w} \rightarrow \infty$ almost surely as $n \to \infty$, completing the proof.
\end{proof}

\begin{proof}{ of Equation \eqref{TS2}}
	Following some algebra, we have from \eqref{ECF} and \eqref{Dn} that
	\begin{align*}
		\big|D^{(n)}(r_1,r_2)\big|^2=\;&\frac{1}{n^2} \sum_{j,k=1}^{n} \cos\big(r_1 \vartheta^{(1)}_{jk}+r_2\vartheta^{(2)}_{jk}\big)+\frac{1}{n^4} \sum_{j,k,\ell,m=1}^n \cos\big(r_1 \vartheta^{(1)}_{jk}+r_2 \vartheta^{(2)}_{\ell m}\big)\nonumber\\ 
		&-\frac{2}{n^3} \sum_{j,k,\ell=1}^{n} \cos\big(r_1 \vartheta^{(1)}_{jk}+r_2\vartheta^{(2)}_{j\ell}\big).
	\end{align*}
	Inserting the above equation in \eqref{TS} we readily obtain 
	\begin{align*}
		T_{n,w}=\frac{1}{n} \sum_{j,k=1}^{n}C_w\big(\vartheta^{(1)}_{jk},\vartheta^{(2)}_{jk}\big)+\frac{1}{n^3} \sum_{j,k,\ell,m=1}^{n}C_w\big(\vartheta^{(1)}_{jk},\vartheta^{(2)}_{\ell m}\big)- \frac{2}{n^2} \sum_{j,k,\ell=1}^{n}C_w\big(\vartheta^{(1)}_{jk},\vartheta^{(2)}_{j\ell}\big),
	\end{align*} 
	where 
	\begin{align}\nonumber
		C_{w}(x,y)=\sum_{r_1=-\infty}^\infty \sum_{r_2=-\infty}^\infty \cos(r_1x+r_2 y) w(r_1,r_2).
	\end{align}
	
	A little reflection shows that, if $w(r_1,r_2)=v(r_1)v(r_2)$ with 
	\begin{align*}
		v(\pm r)=(1/2)f(r), \ r=1,2,\ldots, \quad v(0)=f(0), 
	\end{align*}
	where $f$ is any probability function on the non-negative integers, then the series $C_w(x,y)$ equals ${\cal{J}}^{(f)}_c(x){\cal{J}}^{(f)}_c(y)$ with ${\cal{J}}^{(f)}_c={\cal{J}}^{(v)}_c$ and ${\cal{J}}^{(f)}_c(\vartheta)=\sum_{r=0}^\infty \cos(r\vartheta)f(r)$, which is by definition the real part of the characteristic function corresponding to $f$ evaluated at the point $\vartheta$.
\end{proof}

\section{\texorpdfstring{Derivation of the covariance matrix $\Sigb$}{Derivation of the covariance matrix Sigma}}
\label{sec:Sigma}

The asymptotic normality of $\sqrt{n}{\boldsymbol \Delta}_n\big(\boldsymbol{r}^{(c)},\boldsymbol{r}^{(s)}\big)$, as well as its asymptotic covariance matrix $\Sigb$, follows directly by using the fact that, under a mild tail condition, the empirical characteristic function independence process $\sqrt{n}\left(\hat \varphi(\boldsymbol{r})-\prod_{m=1}^p\hat \varphi_m(r_m)\right)$, $\boldsymbol{r}=(r_1,\ldots,r_m)'\in\mathbb R^p$, $p\geq 1$, converges weakly to a zero-mean complex-valued Gaussian process with covariance structure determined by the matrix $\Sigb$; see \cite{Csorgo1985}. The elements of $\Sigb$ involve the elements of the matrix $\boldsymbol V=(v_{km}(\boldsymbol{r},\boldsymbol{t}))_{k,m=1,2}$, where
\begin{align*}
	{\boldsymbol{V}}=\frac{1}{2}\begin{pmatrix}{\cal{J}}_c(\boldsymbol{r}+\boldsymbol{t})+{\cal{J}}_c(\boldsymbol{r}-\boldsymbol{t})- 2{\cal{J}}_c(\boldsymbol{r}){\cal{J}}_c(\boldsymbol{t}) \ & {\cal{J}}_s(\boldsymbol{r}+\boldsymbol{t})-{\cal{J}}_s(\boldsymbol{r}-\boldsymbol{t})-2 {\cal{J}}_c(\boldsymbol{r}){\cal{J}}_s(\boldsymbol{t}) \\ {\cal{J}}_s(\boldsymbol{r}+\boldsymbol{t})+{\cal{J}}_s(\boldsymbol{r}-\boldsymbol{t})-2 {\cal{J}}_c(\boldsymbol{t}){\cal{J}}_s(\boldsymbol{r}) \ & {\cal{J}}_c(\boldsymbol{r}-\boldsymbol{t})-{\cal{J}}_c(\boldsymbol{r}+\boldsymbol{t})-2 {\cal{J}}_s(\boldsymbol{r}){\cal{J}}_s(\boldsymbol{t}) \end{pmatrix}, 
\end{align*}
with ${\cal{J}}_c(\boldsymbol{r})$ and ${\cal{J}}_s(\boldsymbol{r})$ being the real and imaginary parts, respectively, of $\varphi(\boldsymbol{r})$. The elements of $\Sigb$ ultimately depend on $\varphi$ and $\varphi_m$, $m=1,\ldots,p$. These elements have been obtained by \cite{Csorgo1985} with general dimension $p$, but we report them here for convenience at $p=2$. \\

To this end, define ${\bf{r}}=(r_1,r_2)'$, ${\bf{t}}=(t_1,t_2)'$, ${\bf{r}}_1=(r_1,0)'$, ${\bf{r}}_2=(0,r_2)'$, ${\bf{t}}_1=(t_1,0)'$, and ${\bf{t}}_2=(0,t_2)'$. Also, write ${\cal{J}}_{c_{\neq k}}(\boldsymbol{r})$ and ${\cal{J}}_{s_{\neq k}}(\boldsymbol{r})$ for the real and imaginary parts, respectively, of $\prod_{\substack{m=1\\ m\neq k}}^2\varphi_m(r_m)$. Then, the entries of $\Sigb$ are:
\begin{align*} \label{COV1} \nonumber
	{\rm{cov}}\big(D^{(n)}_c(\boldsymbol{r}),D^{(n)}_c(\boldsymbol{t})\big)=&\;v_{11}(\boldsymbol{r},\boldsymbol{t}) - \sum_{k=1}^2 \Big\{{\cal{J}}_{c_{\neq k}}(\boldsymbol{t})v_{11}(\boldsymbol{r},\boldsymbol{t}_k)-{\cal{J}}_{s_{\neq k}}(\boldsymbol{t})v_{12}(\boldsymbol{r},\boldsymbol{t}_k)\Big\} \\ \nonumber
	&- \sum_{k=1}^2 \Big\{ {\cal{J}}_{c_{\neq k}}(\boldsymbol{r})v_{11}(\boldsymbol{t},\boldsymbol{r}_k)-{\cal{J}}_{s_{\neq k}}(\boldsymbol{r})v_{12}(\boldsymbol{t},\boldsymbol{r}_k)\Big\} \\ \nonumber 
	&+ \sum_{k=1}^2 \sum_{m=1}^2 \Big\{ {\cal{J}}_{c_{\neq k}}(\boldsymbol{r}){\cal{J}}_{c_{\neq m}}(\boldsymbol{t})v_{11}(\boldsymbol{r}_k,\boldsymbol{t}_m)-{\cal{J}}_{c_{\neq k}}(\boldsymbol{r}){\cal{J}}_{s_{\neq m}}(\boldsymbol{t})v_{12}(\boldsymbol{r}_k,\boldsymbol{t}_m)\\\nonumber 
	&-{\cal{J}}_{s_{\neq k}}(\boldsymbol{r}){\cal{J}}_{c_{\neq m}}(\boldsymbol{t})v_{12}(\boldsymbol{t}_m,\boldsymbol{r}_k)+{\cal{J}}_{s_{\neq k}}(\boldsymbol{r}){\cal{J}}_{s_{\neq m}}(\boldsymbol{t})v_{22}(\boldsymbol{r}_k,\boldsymbol{t}_m)\Big\},\\
	{\rm{cov}}\big(D^{(n)}_c(\boldsymbol{r}),D^{(n)}_s(\boldsymbol{t})\big)=&\;v_{12}(\boldsymbol{r},\boldsymbol{t}) - \sum_{k=1}^2 \Big\{{\cal{J}}_{s_{\neq k}}(\boldsymbol{t})v_{11}(\boldsymbol{r},\boldsymbol{t}_k)+{\cal{J}}_{c_{\neq k}}(\boldsymbol{t})v_{12}(\boldsymbol{r},\boldsymbol{t}_k)\Big\} \\ \nonumber
	&- \sum_{k=1}^2 \Big\{ {\cal{J}}_{c_{\neq k}}(\boldsymbol{r})v_{12}(\boldsymbol{r}_k,\boldsymbol{t})-{\cal{J}}_{s_{\neq k}}(\boldsymbol{r})v_{22}(\boldsymbol{r}_k,\boldsymbol{t})\Big\} \\ \nonumber 
	&+ \sum_{k=1}^2 \sum_{m=1}^2 \Big\{ {\cal{J}}_{c_{\neq k}}(\boldsymbol{r}){\cal{J}}_{s_{\neq m}}(\boldsymbol{t})v_{11}(\boldsymbol{r}_k,\boldsymbol{t}_m)+{\cal{J}}_{c_{\neq k}}(\boldsymbol{r}){\cal{J}}_{c_{\neq m}}(\boldsymbol{t})v_{12}(\boldsymbol{r}_k,\boldsymbol{t}_m)\\ \nonumber 
	&-{\cal{J}}_{s_{\neq k}}(\boldsymbol{r}){\cal{J}}_{s_{\neq m}}(\boldsymbol{t})v_{12}(\boldsymbol{t}_m,\boldsymbol{r}_k)-{\cal{J}}_{s_{\neq k}}(\boldsymbol{r}){\cal{J}}_{c_{\neq m}}(\boldsymbol{t})v_{22}(\boldsymbol{r}_k,\boldsymbol{t}_m)\Big\},\\
	{\rm{cov}}\big(D^{(n)}_s(\boldsymbol{r}),D^{(n)}_s(\boldsymbol{t})\big)=&\;v_{22}(\boldsymbol{r},\boldsymbol{t}) - \sum_{k=1}^2 \Big\{{\cal{J}}_{s_{\neq k}}(\boldsymbol{t})v_{12}(\boldsymbol{t}_k,\boldsymbol{r})+{\cal{J}}_{c_{\neq k}}(\boldsymbol{t})v_{22}(\boldsymbol{r},\boldsymbol{t}_k)\Big\} \\ \nonumber
	&- \sum_{k=1}^2 \Big\{ {\cal{J}}_{s_{\neq k}}(\boldsymbol{r})v_{12}(\boldsymbol{r}_k,\boldsymbol{t})+{\cal{J}}_{c_{\neq k}}(\boldsymbol{r})v_{22}(\boldsymbol{r}_k,\boldsymbol{t})\Big\} \\ \nonumber 
	&+ \sum_{k=1}^2 \sum_{m=1}^2 \Big\{ {\cal{J}}_{s_{\neq k}}(\boldsymbol{r}){\cal{J}}_{s_{\neq m}}(\boldsymbol{t})v_{11}(\boldsymbol{r}_k,\boldsymbol{t}_m)+{\cal{J}}_{s_{\neq k}}(\boldsymbol{r}){\cal{J}}_{c_{\neq m}}(\boldsymbol{t})v_{12}(\boldsymbol{r}_k,\boldsymbol{t}_m)\\ \nonumber 
	&+{\cal{J}}_{c_{\neq k}}(\boldsymbol{r}){\cal{J}}_{s_{\neq m}}(\boldsymbol{t})v_{12}(\boldsymbol{t}_m,\boldsymbol{r}_k)+{\cal{J}}_{c_{\neq k}}(\boldsymbol{r}){\cal{J}}_{c_{\neq m}}(\boldsymbol{t})v_{22}(\boldsymbol{r}_k,\boldsymbol{t}_m)\Big\}.
\end{align*}

\section{Permutation approach}
\label{sec:perm}

The following permutation algorithm can be employed to emit a decision in $\phi\n(\lambda)$. The algorithm is standard, yet it highlights the aspects to produce a computationally efficient permutation approach tailored for the test statistic $T_{n,\lambda}$.

\begin{algo}\label{algo}
	Assume $\big(\vartheta_1^{(1)}, \vartheta_1^{(2)}\big), \ldots, \big(\vartheta_n^{(1)}, \vartheta_n^{(2)}\big)$ and $\lambda$ are given.
	\begin{enumerate}
		\item Compute ${\cal{J}}^{(v)}_c\big(\vartheta^{(m)}_{jk}\big)$, $j,k=1,\ldots,n$, $m=1,2$, using \eqref{poisson}.
		\item Compute $T_{n,\lambda}$ using \eqref{TS2}. Recall that the third term is the sum of a matrix multiplication.
		\item For $b=1,\ldots,B$:
		\begin{enumerate}
			\item Sample without replacement $\vartheta_1^{(2,*b)},\ldots,\vartheta_n^{(2,*b)}$ from $\big\{\vartheta_1^{(2)},\ldots,\vartheta_n^{(2)}\big\}$.
			\item Compute ${\cal{J}}^{(v)}_c\big(\vartheta^{(2,*b)}_{jk}\big)$, $j,k=1,\ldots,n$, using \eqref{poisson}.
			\item Compute $T_{n,\lambda}^{(*b)}$ from ${\cal{J}}^{(v)}_c\big(\vartheta^{(1)}_{jk}\big)$ and ${\cal{J}}^{(v)}_c\big(\vartheta^{(2,*b)}_{jk}\big)$, $j,k=1,\ldots,n$, using \eqref{TS2}. Recall that the second term in $T_{n,\lambda}$ is the same as in $T_{n,\lambda}^{(*b)}$, hence it can be saved.
		\end{enumerate}
		\item Set the permutation-approximated $p$-value as $B^{-1}\sum_{b=1}^B1_{\big\{T_{n,\lambda}<T^{(*b)}_{n,\lambda}\big\}}$.
	\end{enumerate}
\end{algo}

Analogous permutation approaches can be set for the $\phi_c^{(n)}(r_1,r_2)$ and $\phi\n\big(\boldsymbol{r}^{(c)},\boldsymbol{r}^{(s)}\big)$ tests, although these are not practically needed thanks to their usable null asymptotic distributions.

\section{Further simulation results}
\label{sec:moresimus}

In a Monte Carlo simulation setting, computing critical thresholds for all test statistics can be done either in a nonparametric way (using $B$ permutations) or in a purely parametric way (with $M_c$ resamples) as explained in Section \ref{subsec:empiricalpowers}. The latter approach is much faster, and we claim that it can be used in our context to conduct Monte Carlo experiments in order to compare the power of the ten tests under scrutiny. To backup this claim, we investigate the proximity of the empirical power obtained (with $M$ Monte Carlo resamples) using the parametric approach to the one using a nonparametric approach based on permutations. The sample sizes $n=20$ and $n=50$ are considered, and we set the significance level to the value $\alpha=0.05$. We added to each empirical power point a (classical) binomial confidence interval with a Bonferroni-corrected confidence level set to $1 - 0.05 / K$, where $K$ is the number of confidence intervals on a given plot. Figure~\ref{fig:scenario1-2} illustrates the results for the scenarios of dependence \ref{mod1} and \ref{mod2}, while Figure~\ref{fig:scenario3-4} illustrates the results for \ref{mod3} and \ref{mod4}.

\vspace*{-0.1cm}

\begin{figure}[h]
	\centering
	\includegraphics[width=15cm]{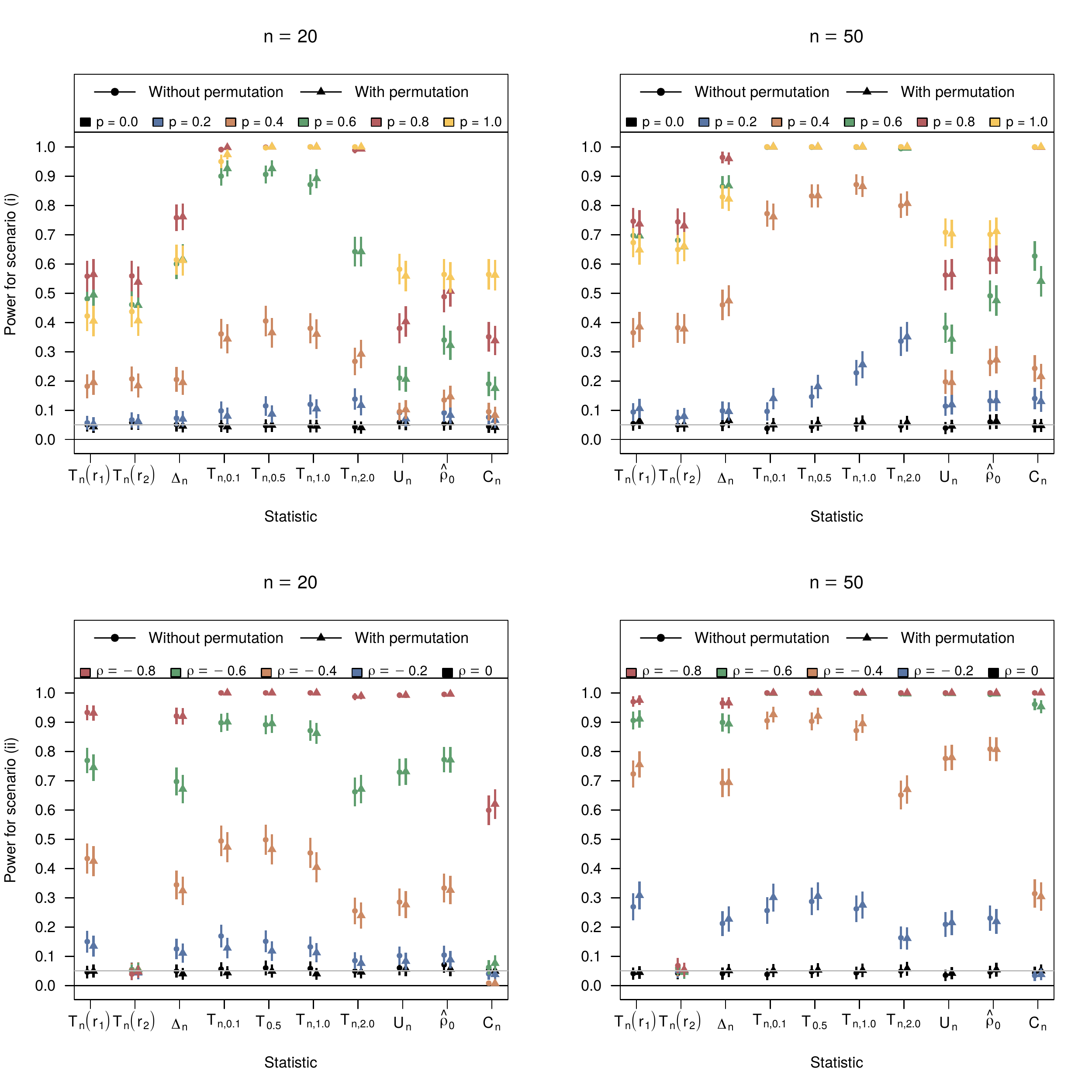}
	\caption{\small Comparison of the approach using permutations ($M=B=10^3$) and the one without ($M=M_c=10^3)$ for scenarios \ref{mod1} (top row) and \ref{mod2} (bottom row).}\label{fig:scenario1-2}
\end{figure}

The conclusion of this study is that the overall pattern of all power values is very similar for both approaches. This justifies our choice to use the parametric approach to create Table~\ref{tab:4.5.1} for $n=50$, the equivalent of which for $n=20$ being Table \ref{tab:4.5.2} below. It is worth mentioning, however, that some minor discrepancies can be observed. The most notable is in scenario \ref{mod3} when $n=20$, for which we observe slightly higher values of power for $T_{n,\lambda}$ using the permutation approach (without changing the power ranking). For small sample sizes and small significance levels (not shown here), we sometimes also observed higher values of power for $T_{n,\lambda}$ using the permutation approach.\\

Finally, to get a better grasp of how dependence is controlled by adjusting the value of the parameter $p$, $\rho$, $\kappa_3$, or $\kappa_g$, we illustrate in Figure~\ref{fig:2} below the resulting shapes of dependence for a few simulated samples.

\begin{figure}[H]
	\centering
	\includegraphics[width=15cm]{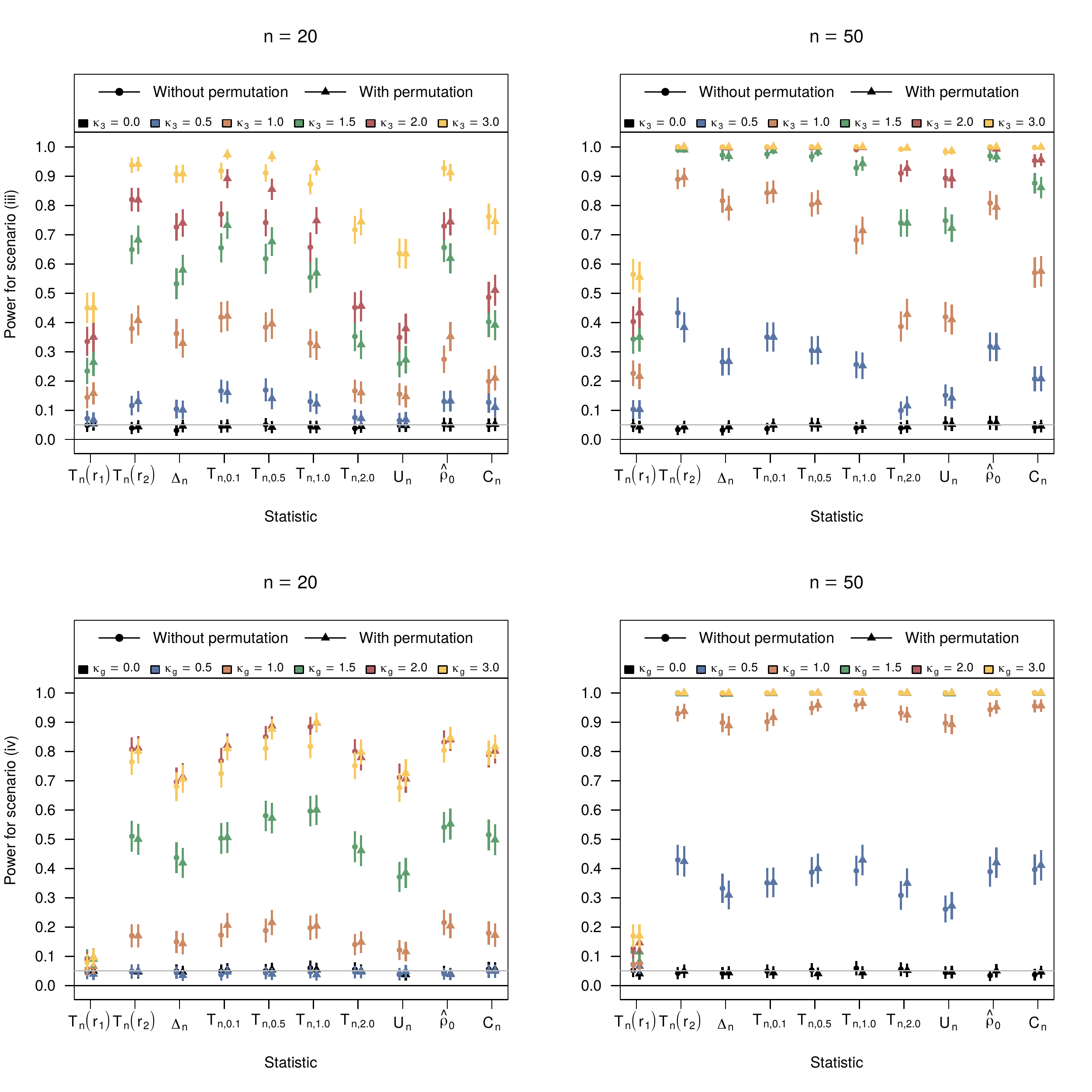}
	\caption{\small Comparison of the approach using permutations ($M=B=10^3$) and the one without ($M=M_c=10^3)$ for scenarios \ref{mod3} (top row) and \ref{mod4} (bottom row).}\label{fig:scenario3-4}
\end{figure}

\begin{table}[h]
	\centering
	\footnotesize
		\begin{tabular}{cc|rrrrrrrrrr}
			\toprule
			&     & $T_n(\boldsymbol{r}_1)$ & $T_n(\boldsymbol{r}_2)$ & $\boldsymbol{\Delta}_n$ & $T_{n,0.1}$ & $T_{n,0.5}$ & $T_{n,1.0}$ & $T_{n,2.0}$ & $U_n$ & $\hat{\rho}_0$ & $C_n$ \\
			\midrule
			\multirow{6}{*}{$p$} & 0.0 & 4.99 & 4.99 & 5.13 & 5.01 & 4.99 & 4.93 & 5.11 & 4.79 & 5.02 & 4.94 \\
			& 0.2 & 6.68 & 6.49 & 7.14 & 7.42 & 8.37 & 10.25 & \textbf{12.09} & 7.27 & 8.04 & 7.19 \\
			& 0.4 & 18.50 & 18.28 & 20.72 & 32.39 & 34.71 & \textbf{35.87} & 28.99 & 10.01 & 13.83 & 9.40 \\
			& 0.6 & 46.70 & 46.77 & 58.21 & 86.43 & \textbf{86.75} & 84.26 & 65.05 & 18.11 & 28.68 & 16.16 \\
			& 0.8 & 53.88 & 53.62 & 74.36 & 98.04 & 99.22 & \textbf{99.54} & 98.84 & 37.05 & 45.28 & 32.35 \\
			& 1.0 & 41.54 & 41.35 & 61.97 & 94.12 & 99.48 & 99.96 & \textbf{100.00} & 55.02 & 54.65 & 56.17 \\
			\midrule
			\multirow{5}{*}{$\rho$} & 0.0 & 5.06 & 5.04 & 4.97 & 5.11 & 4.97 & 4.94 & 4.80 & 4.85 & 5.00 & 4.92 \\
			& 0.2 & \textbf{13.94} & 5.08 & 10.91 & \textbf{13.88} & 13.50 & 11.95 & 8.04 & 8.90 & 10.08 & 3.71 \\
			& 0.4 & 42.81 & 5.14 & 33.72 & \textbf{48.40} & 47.10 & 42.40 & 24.22 & 30.95 & 34.37 & 1.70 \\
			& 0.6 & 75.52 & 5.40 & 67.99 & \textbf{88.62} & 88.32 & 85.85 & 66.88 & 74.61 & 78.44 & 7.97 \\
			& 0.8 & 93.49 & 5.41 & 91.77 & 99.70 & \textbf{99.76} & \textbf{99.73} & 98.75 & 98.92 & 99.23 & 63.36 \\
			\midrule
			\multirow{6}{*}{$\kappa_3$} & 0.0 & 4.96 & 5.00 & 4.98 & 5.07 & 5.02 & 5.13 & 5.17 & 4.93 & 5.17 & 4.93 \\
			& 0.5 & 8.37 & 14.73 & 12.03 & \textbf{15.91} & 14.80 & 12.45 & 8.22 & 7.11 & 12.83 & 9.19 \\
			& 1.0 & 15.83 & \textbf{42.41} & 33.43 & 40.99 & 38.04 & 31.60 & 17.11 & 14.75 & 33.84 & 21.35 \\
			& 1.5 & 25.56 & \textbf{68.50} & 57.27 & 64.40 & 61.17 & 52.90 & 30.46 & 25.84 & 58.30 & 36.97 \\
			& 2.0 & 34.52 & \textbf{83.81} & 75.05 & 79.35 & 77.01 & 70.10 & 45.46 & 40.13 & 76.49 & 52.22 \\
			& 3.0 & 45.70 & \textbf{94.54} & 90.66 & 91.02 & 90.39 & 87.29 & 70.48 & 61.36 & 93.10 & 73.55 \\
			\midrule
			\multirow{6}{*}{$\kappa_g$} & 0.0 & 4.98 & 5.07 & 4.96 & 5.01 & 4.93 & 5.07 & 5.10 & 4.82 & 5.07 & 5.04 \\
			& 0.5 & 6.73 & 16.37 & 13.24 & 16.13 & \textbf{17.66} & \textbf{17.81} & 13.76 & 11.19 & \textbf{17.89} & 16.41 \\
			& 1.0 & 8.77 & 50.11 & 40.32 & 45.93 & 51.95 & \textbf{54.84} & 43.87 & 37.74 & 53.20 & 49.49 \\
			& 1.5 & 10.59 & 78.89 & 71.06 & 76.13 & 83.93 & \textbf{86.82} & 76.90 & 71.22 & 83.01 & 81.28 \\
			& 2.0 & 12.14 & 91.82 & 87.86 & 91.35 & 96.05 & \textbf{97.54} & 93.57 & 90.18 & 95.36 & 94.92 \\
			& 3.0 & 14.06 & 98.19 & 97.46 & 98.70 & 99.65 & \textbf{99.87} & 99.63 & 98.94 & 99.65 & 99.62 \\
			\bottomrule
		\end{tabular}
	\caption{\small Empirical level and power (in \%) for the distributions $\mathrm{PB}(p)$, $\mathrm{BWC}(0.1,0.1,-\rho)$, $\mathrm{BCvM}(1,1,\kappa_3)$, and $\mathrm{BvM}(1, 1, 0, \kappa_g)$ (top to bottom), for $\alpha=5\%$ and $n=20$. On each row, the largest power value is in bold, and any other power value falling in the \cite{Wilson1927}'s $95\%$ binomial confidence interval for the theoretical power of this best test is also in bold.}\label{tab:4.5.2}
\end{table}

\begin{figure}[h]
	\centering
	\includegraphics[width=0.25\textwidth]{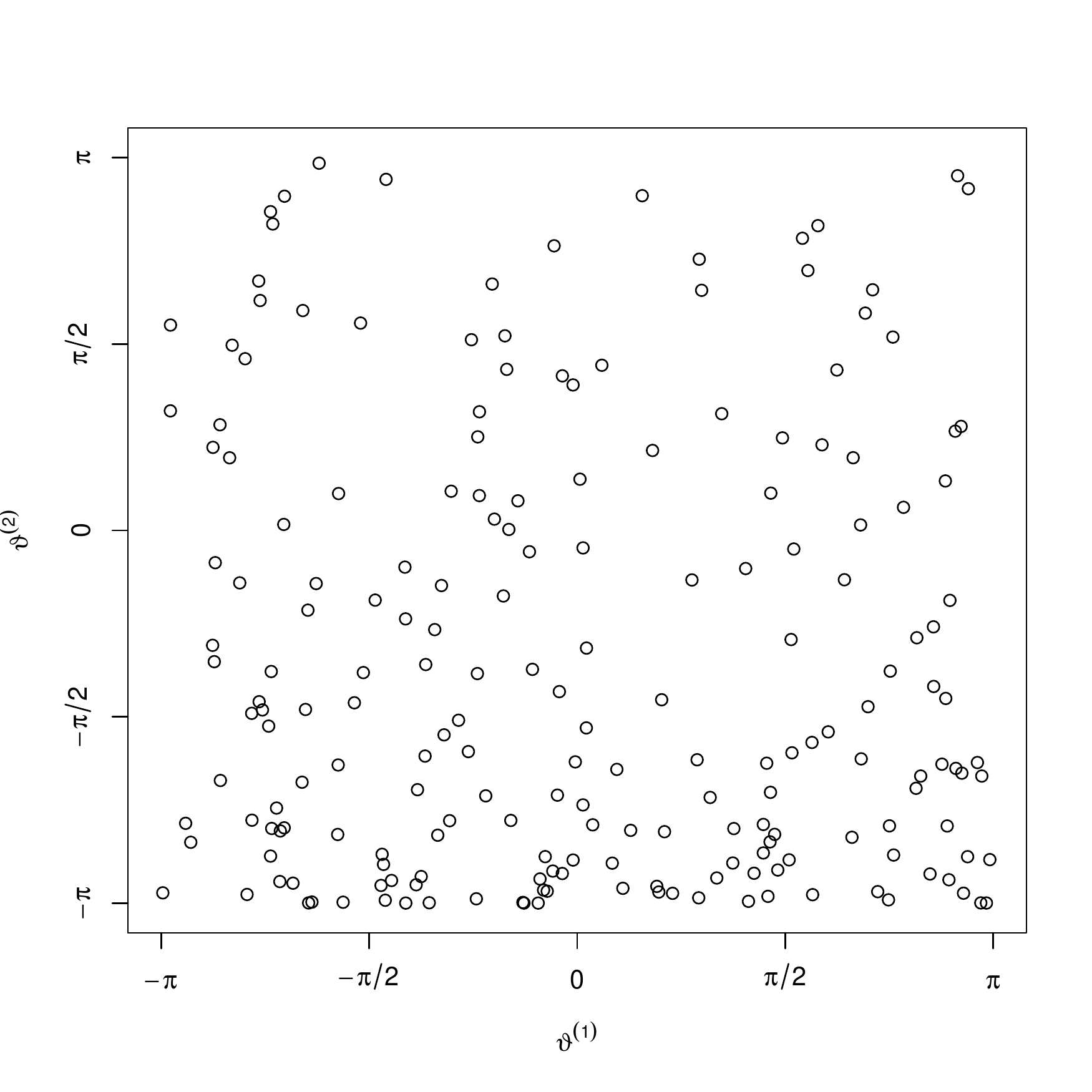}\includegraphics[width=0.25\textwidth]{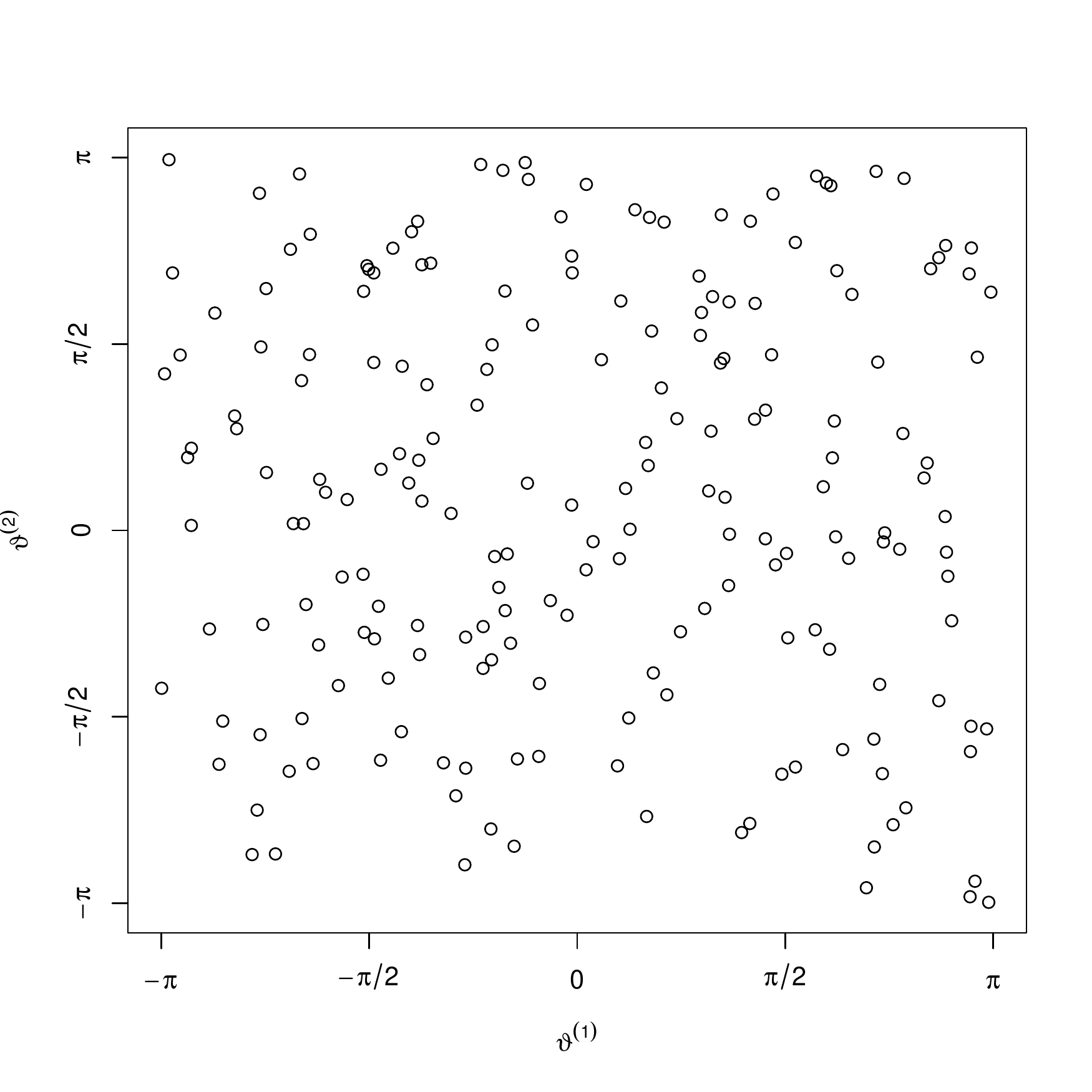}%
	\includegraphics[width=0.25\textwidth]{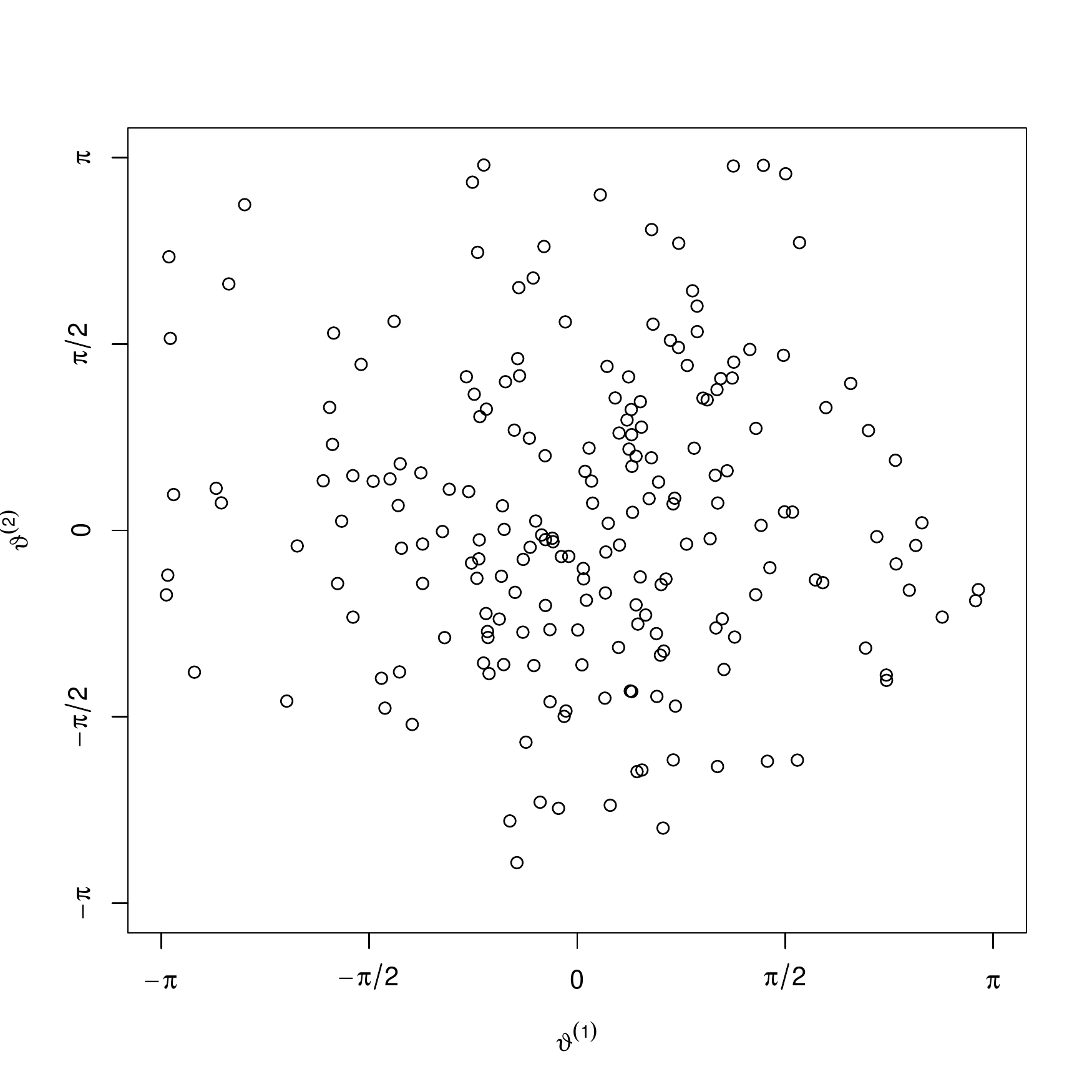}\includegraphics[width=0.25\textwidth]{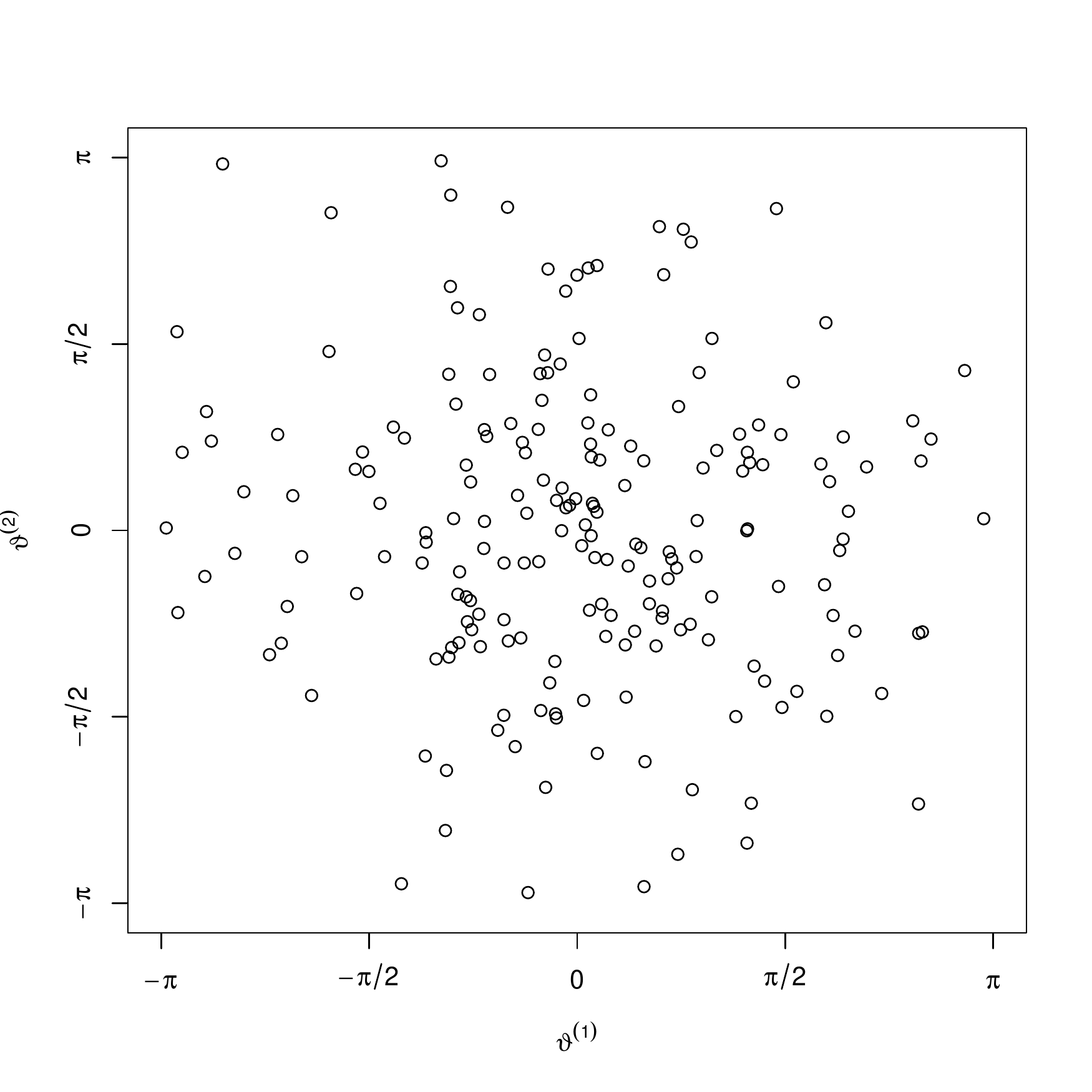}\\
	\includegraphics[width=0.25\textwidth]{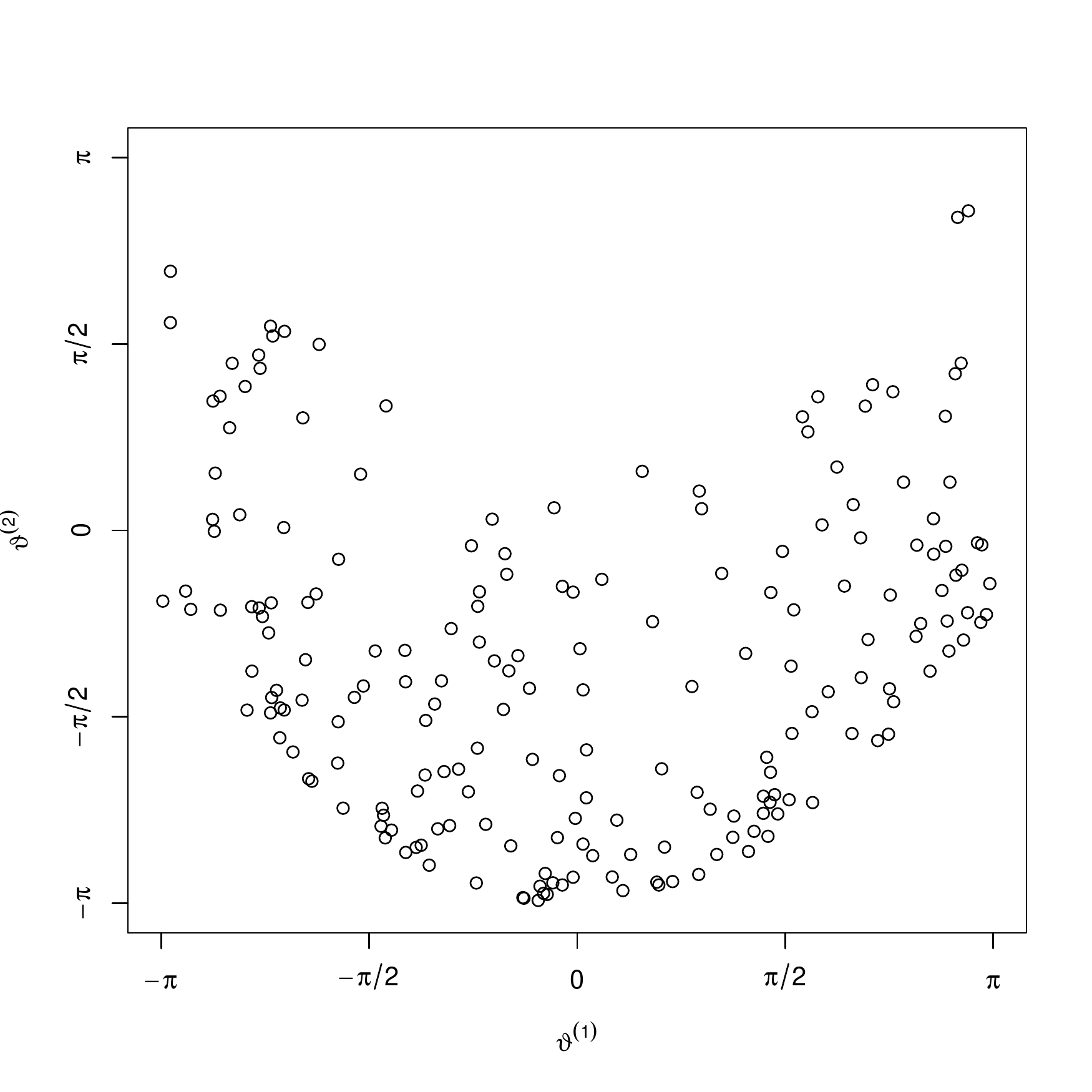}\includegraphics[width=0.25\textwidth]{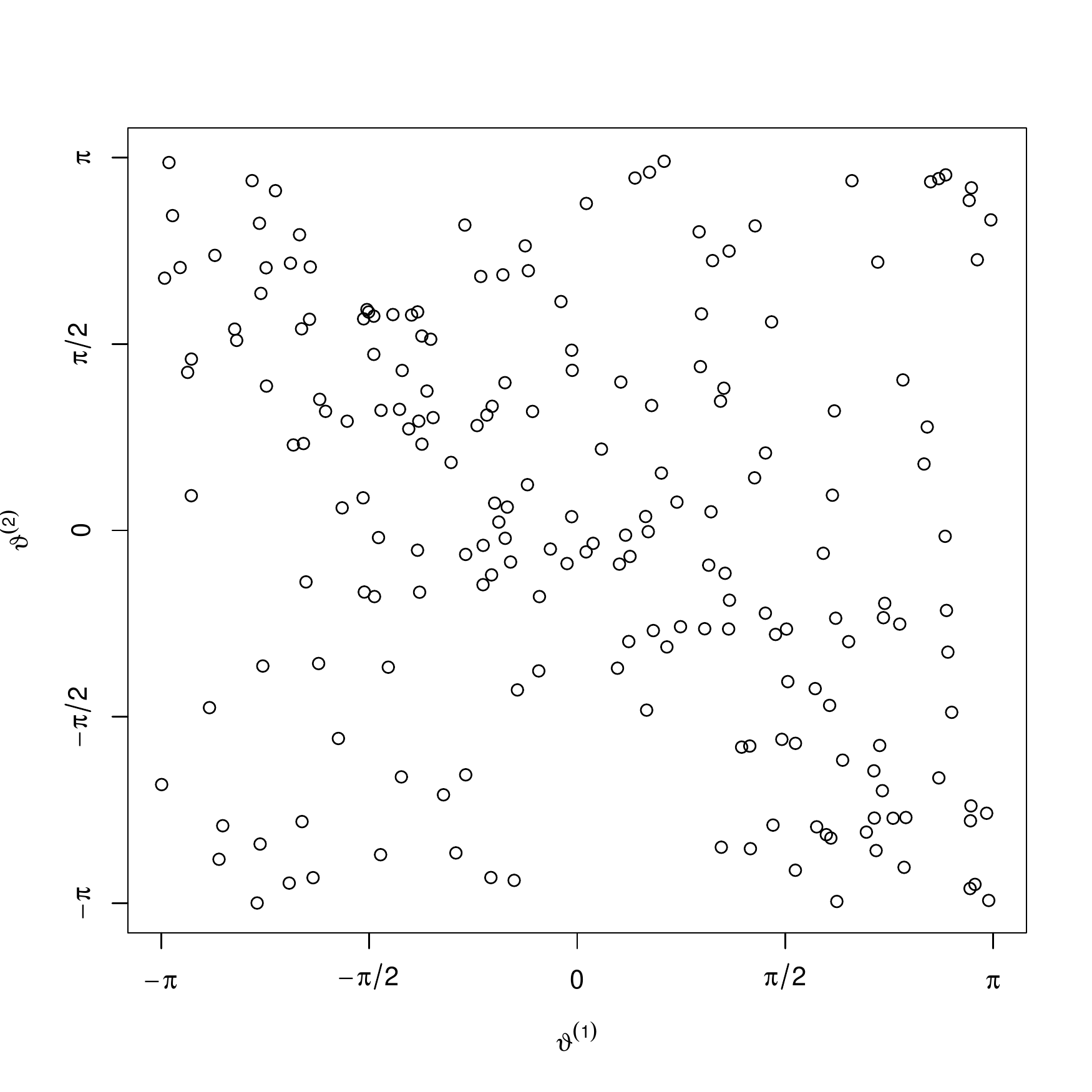}%
	\includegraphics[width=0.25\textwidth]{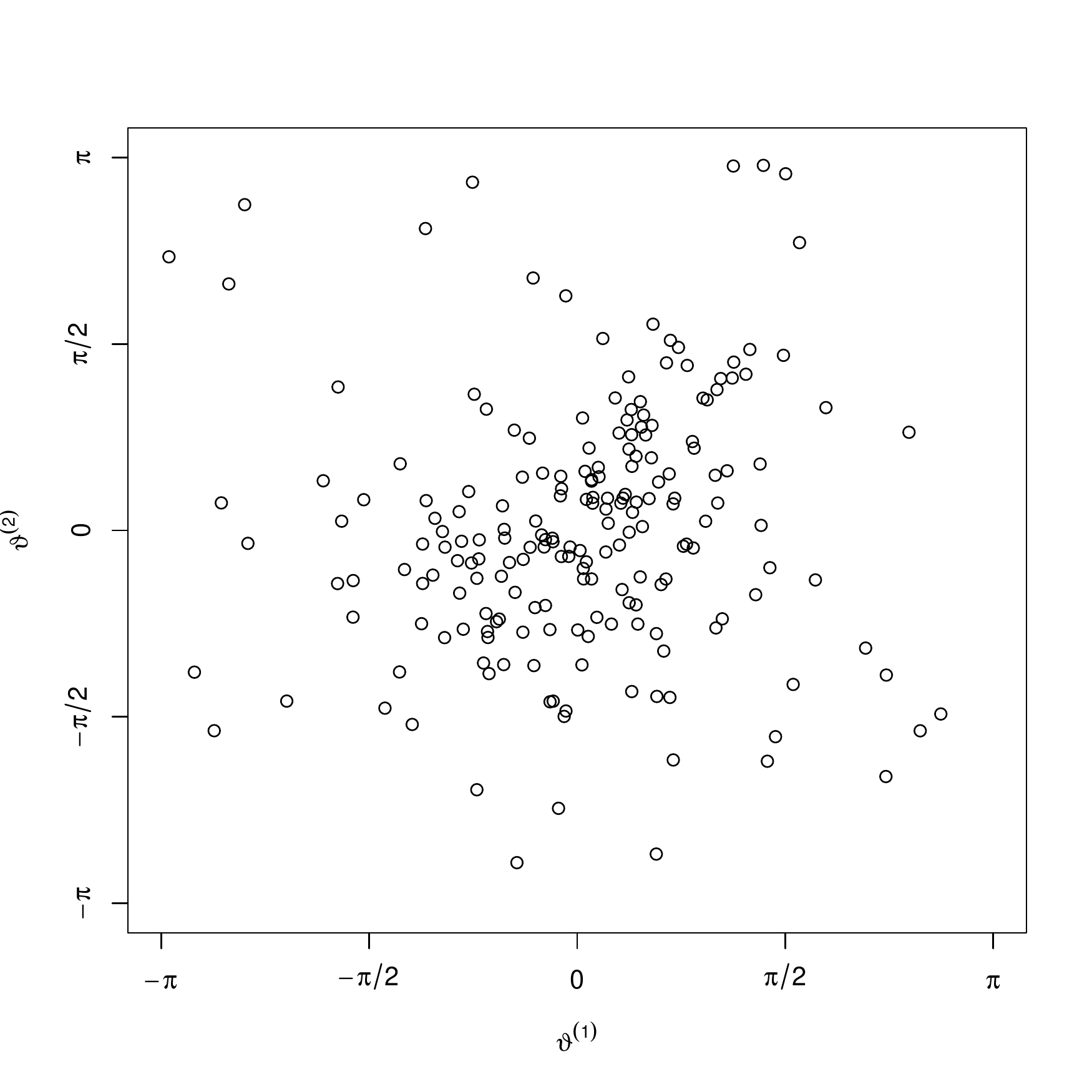}\includegraphics[width=0.25\textwidth]{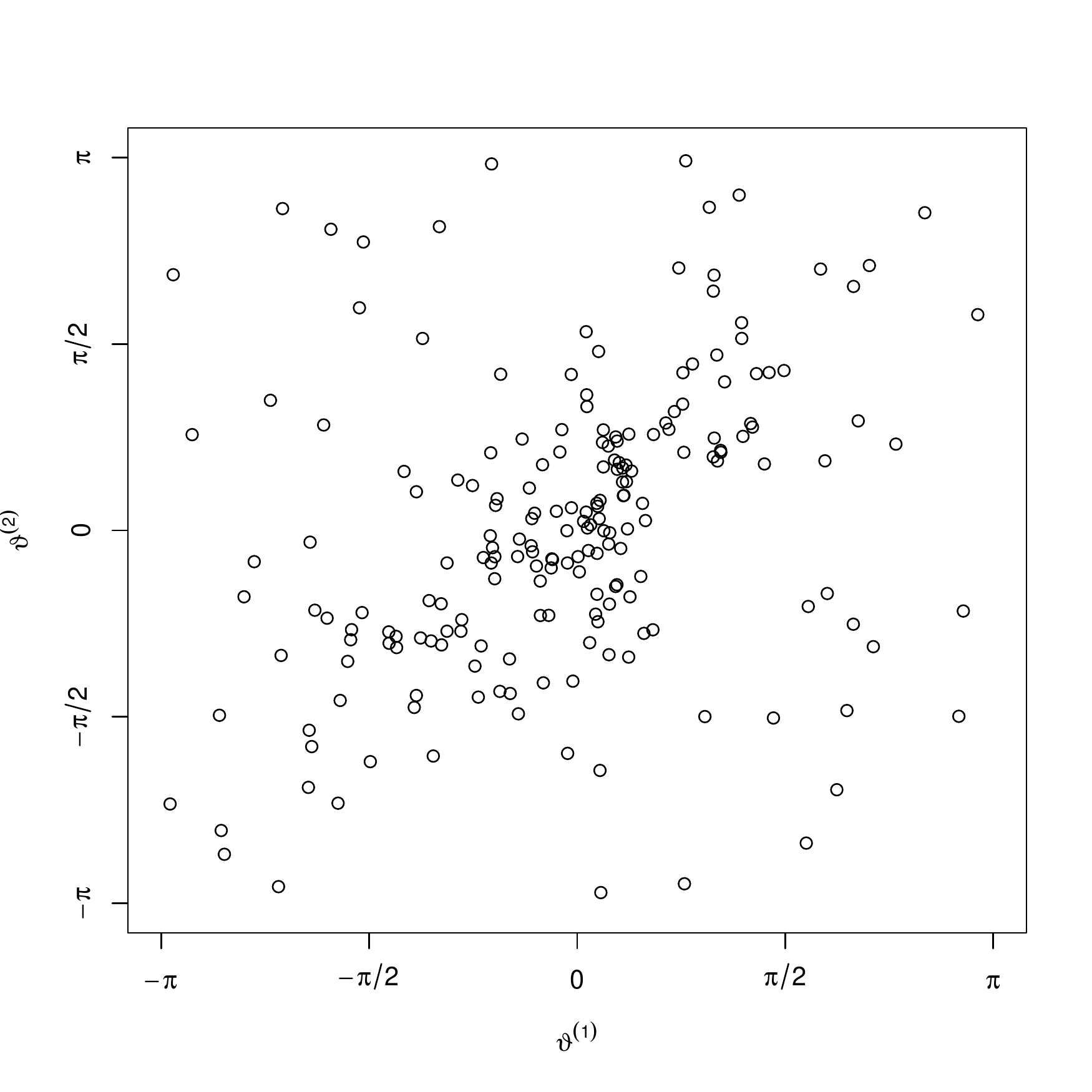}\\
	\includegraphics[width=0.25\textwidth]{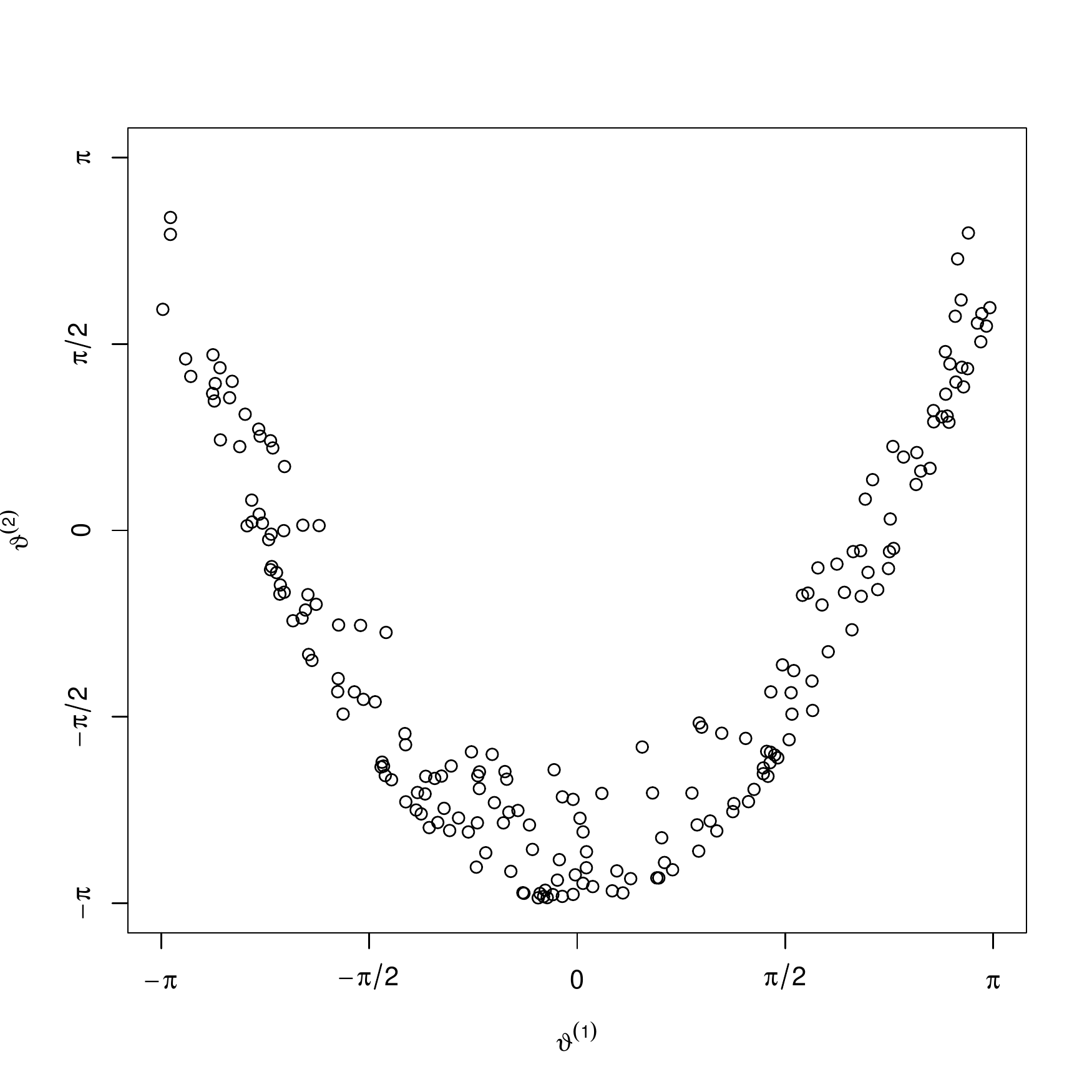}\includegraphics[width=0.25\textwidth]{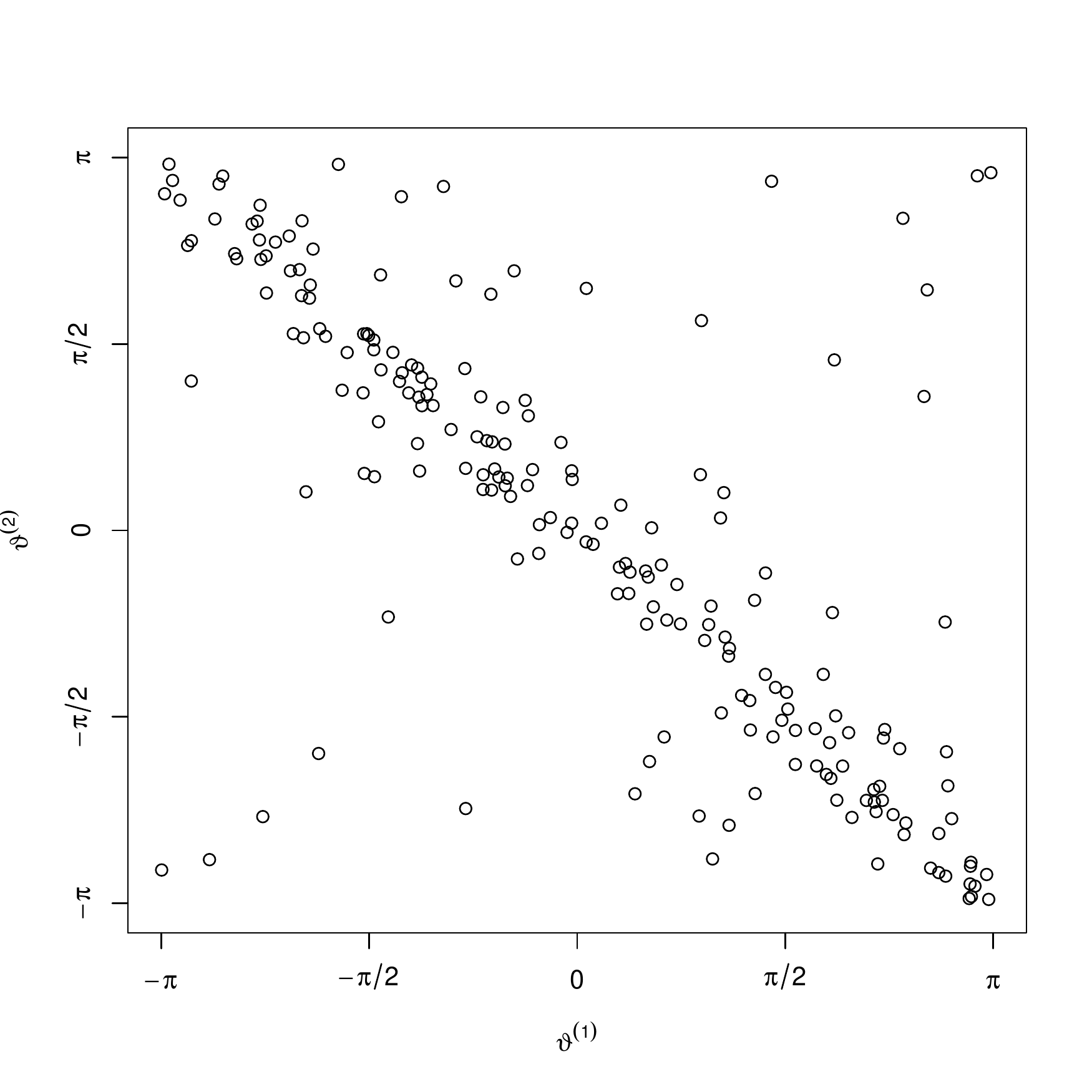}%
	\includegraphics[width=0.25\textwidth]{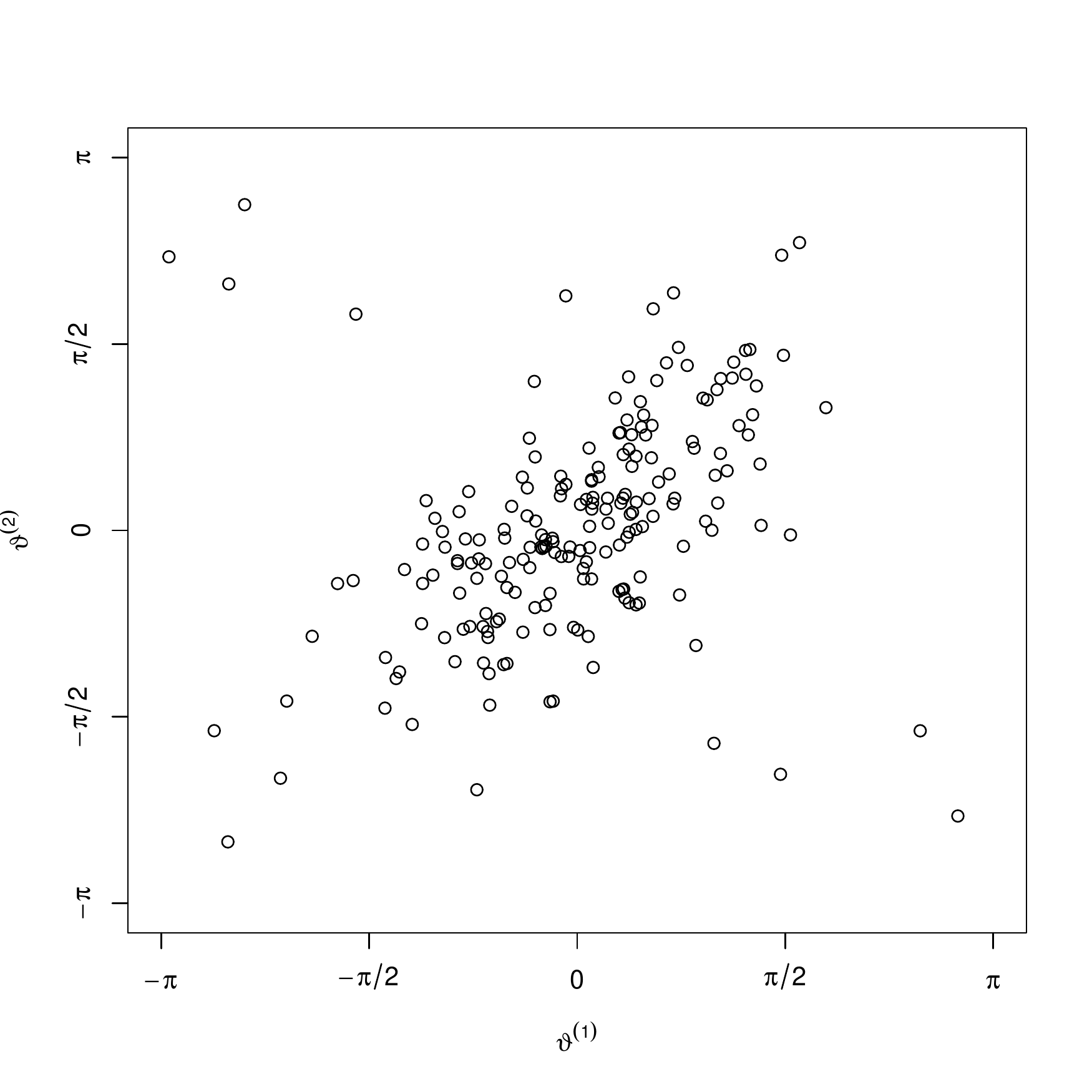}\includegraphics[width=0.25\textwidth]{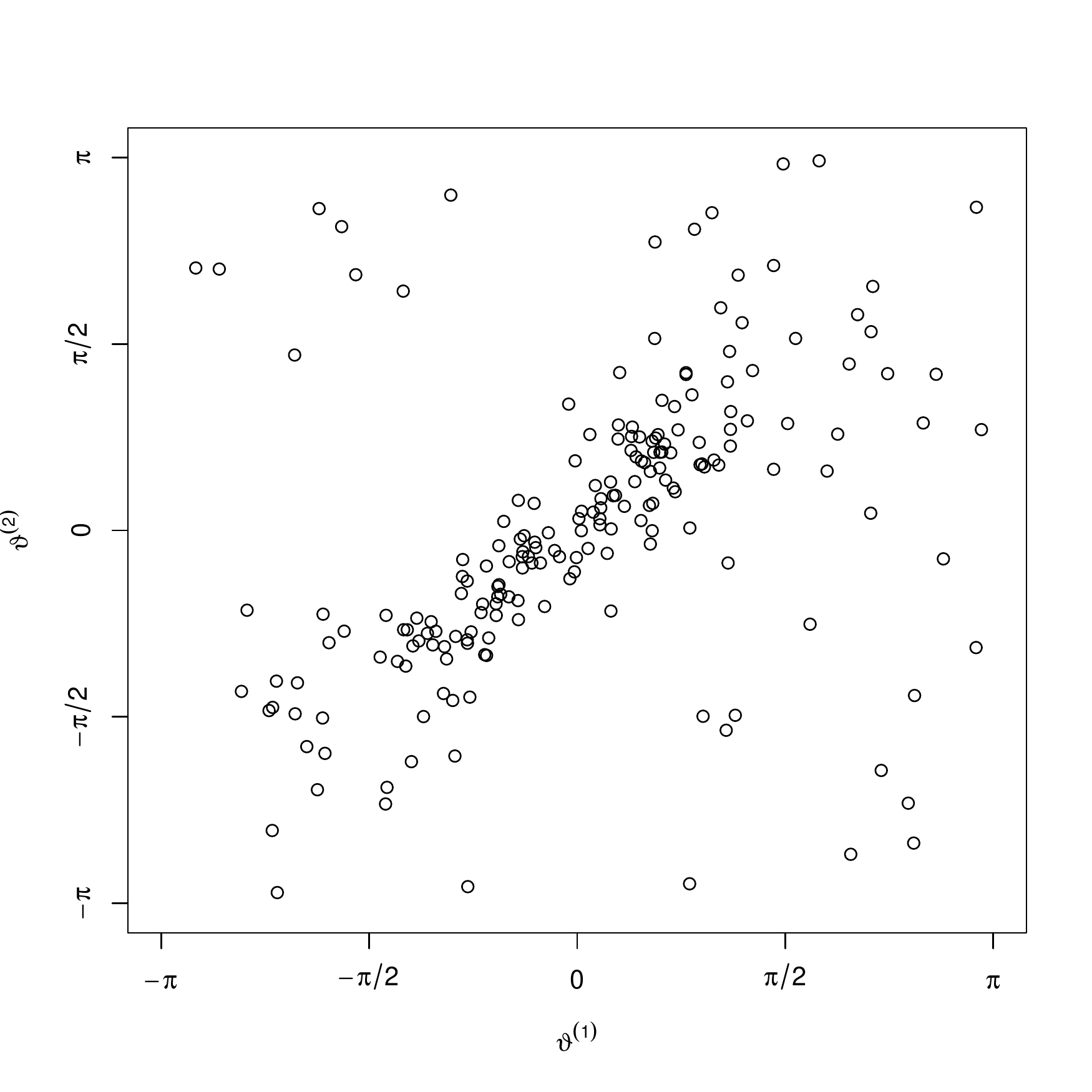}
	\caption{\small Scatterplots generated from the simulation scenarios considered in the simulation study. From left to right, columnwise: \ref{mod1} $\mathrm{PB}(p)$ for $p=0,0.4,0.8$ (top to bottom); \ref{mod2} $\mathrm{BWC}(0.1, 0.1, -\rho)$ for $\rho=0,0.4,0.8$; \ref{mod3} $\mathrm{BCvM}(1, 1, \kappa_3)$ for $\kappa_3=0,1,2$; \ref{mod4} $\mathrm{BvM}(1, 1, 0, \kappa_g)$ for $\kappa_g=0,1,2$. The sample size considered is $n=200$. \label{fig:2}}
\end{figure}


\fi

\end{document}